\documentclass[aps,pra,twocolumn,showpacs,nofootinbib,longbibliography,notitlepage]{revtex4-2}
\usepackage{etex}
\usepackage{amsmath,amssymb,amsthm}
\usepackage[colorlinks=true,citecolor=blue,urlcolor=blue]{hyperref}
\usepackage[pdftex]{graphicx}
\usepackage{times,txfonts}
\usepackage{braket}
\usepackage{color}
\usepackage{natbib}
\usepackage{amsmath,blkarray}
\usepackage{mathtools}
\usepackage{latexsym}
\usepackage{tabularx, booktabs}
\usepackage{graphics,epstopdf}
\usepackage{graphicx}
\usepackage{float}
\usepackage{graphicx}
\usepackage{amsfonts}
\usepackage{subcaption}
\usepackage{color,soul}

\newcommand{\be}{\begin{equation}}
	\newcommand{\ee}{\end{equation}}
\newcommand{\ba}{\begin{eqnarray}}
	\newcommand{\ea}{\end{eqnarray}}

\usepackage{romannum}
\usepackage{bbold}
\usepackage{multirow}

\makeatletter
\newcommand*{\rom}[1]{\expandafter\@slowromancap\romannumeral #1@}
\makeatother

\begin{document}
\title{Tensile quantum-to-classical transition of
macroscopic entangled states under complete
coarse-grained measurements }
\author{ Laxmi Prasad Naik }
\email{laxmiprasadnaik@5897@gmail.com}
\author{ Tamal Ghosh }
\email{tamalghosh695@gmail.com}
\author{ Sumit Mukherjee }
\email{mukherjeesumit93@gmail.com}
\author{ Chiranjib Mitra }
\email{chiranjib@iiserkol.ac.in}
\author{ Prasanta K. Panigrahi }
\email{pprasanta@iiserkol.ac.in}
\affiliation{Department of Physical Sciences, Indian Institute of Science Education and Research Kolkata, Mohanpur 741246, West Bengal, India}
\begin{abstract}
  The departure of quantum world from classical regime is captured through the observation of non-classical correlations manifested in the behaviors of subatomic systems. However, once the dimension of the system becomes substantially large, the quantum behavior begins to decline and subsequently, it starts following the predictions of classical physics. The macroscopic limit at which such quantum-to-classical transition occurs remains one of the long-standing questions in the foundations of quantum theory. There are evidences that the macroscopic limit to which the \textit{quantumness} of a system persists depends on the degree of interaction due to the measurement processes. For instance, with a system having a considerably large Hilbert space dimension, if the measurement is performed in such a way that the outcome of the measurement only reveals a coarse-grained version of the information about the individual level of the concerned system then the disturbance due to the measurement process can be considered to be infinitesimally small. Based on such coarse-grained measurement the dependence of Bell inequality violation on the degree of coarsening has already been investigated \textcolor{blue}{[Phys. Rev. Lett. 112, 010402 (2014)]}. In this paper, we first capture the fact that when local-realism is taken to be the defining notion of classicality, the effect of the degree of coarsening on the downfall of \textit{quantumness} of a macroscopic entangled state can be compensated by testing a Bell-inequality of a higher number of settings from a family of symmetric Bell-inequalities if the number of settings is odd. However, on the contrary, we show that such compensation can not be seen when we witness such quantum-to-classical transition using symmetric Bell inequalities having an even number of settings. Finally, complementing the above result, we show that when unsteerability is taken as the classicality, for both odd and even numbers of settings the degree of coarsening at which the quantum-to-classical transition occurs can be consistently pushed ahead by testing a linear steering inequality of a higher number of settings and observing its violation. We further extend our treatment for mixed macroscopic entangled states.
\end{abstract}
\maketitle
\section{Introduction}
Quantum theory provides a set of mathematical tools that accurately predict most of the characteristic features of microscopic entities that emerge by means of experimental statistics. However, the predictions of quantum theory for certain phenomenology hinder any explanation using \textit{classical} laws that are constructed based on our observations about the behavior of the macroscopic world. Surprisingly, in spite of the fact, that the postulates of quantum theory do not provide any restrictions on its applicability based on the degree of macroscopic dimension of systems, it is seen that as the dimension becomes large up to a certain degree, quantum phenomena eventually breaks down and systems start behaving classically. Although, there are explanations of such quantum-to-classical transition of a system, viz., decoherence program \cite{W.H.Zurek}, collapse models \cite{A.Bassi}, coarse-grained measurements \cite{H.Jeong,
J.Kofler}, still it is not properly understood where the quantum-to-classical boundary lies if there is one. In order to find and estimate the factors that contribute to such quantum-to-classical transition, one has to first define precisely the notion of \textit{classicality} relevant to a particular context. We discuss these notions of \textit{classicality} relevant to the current paper in the following paragraph.

Typically, any particular notion of classicality is formulated based on certain assumptions at the level of preparation, transformation, or measurement related to either single or composite systems. In contrast to the notions of classicality such as macroscopic realism \cite{A.J.Leggett1985,A.J.Leggett2002,C.Emary}, and noncontextuality \cite{R.W.Spekkens,KSpeker68}, based on assumptions about single-particle correlations, classicality for composite systems essentially consists of assumptions about global correlations. In this paper, we mainly consider classical correlations based on spatially separated bipartite systems, for which the most important notions of classicality are local-realism \cite{J.S.Bell, N.Brunner} and unsteerability \cite{H.M.Wiseman,R.Uola, S.J.Jones}. The particular feature for which certain quantum correlations do not satisfy local realism and unsteerability is commonly termed nonlocality and steering respectively. It is important at this point to note, that the degree to which one can observe nonclassicality in a bipartite system, depends on how and in what accuracy the measurements are performed on it. Although there have been studies \cite{asmita2019,silva2015,sasmal2018,zhang21,brown2020} on how the unsharp measurements \cite{P.Busch} affect the bipartite correlation, there are only a few works that considered limitations in measurement accuracy in multilevel systems, in terms of what is called the coarse-grained measurements \cite{J.Kofler}.

In quantum theory, the measurement of an observable is represented through positive semi-definite operators. For a multilevel system, a measurement is expected to reveal a value of the concerned observable, corresponding to any of the levels, where the system has a nonzero probability of being found. For instance, a spin-measurement on a $n$-level spin-system should have $n$ distinct values. However, due to the limited resolution of the measurement devices, it is possible that the measurement outcome only reveals information about a coarse-grained version of individual levels of the system. The effect of such measurement imprecision on the dynamics of a quantum system has been well studied in literature \cite{K.V.Hovhannisyan,P.P.Potts, A.Garg, V.Krishnan, C.Roh}. In reference to the particular notion of classicality known as macrorealism, captured by the Leggett-Garg inequalities \cite{A.J.Leggett1985}, Kofler \textit{et. al.} studied \cite{J.Kofler} the role of measurement imprecision in quantum-to-classical transitions. Moreover, such studies are extended for multilevel spin systems, by taking into account other witnesses \cite{S.Mal, Y.Zhang22} of macrorealism as well as incorporating additional constraints \cite{Y.Zhang23}. Recently, the effect of experimental imprecision other than that of coarsening of measurements on the amount of bipartite quantum correlations has also been investigated in \cite{P.Roy,roh1,roh2}.

Against the backdrop, it is observed \cite{H.Jeong, S.Mukherjee} that in addition to the imprecision in final measurement resolution, the imperfection in the measurement reference also plays an important role in the quantum-to-classical transition. In particular, the ref.\cite{H.Jeong}, considered bipartite nonlocal correlations as a witness of nonclassicality (that are captured through the violation of Bell-CHSH inequality \cite{J.S.Bell,J.F.Clauser}), and demonstrated the quantum-to-classical transition of a macroscopic entangled state under the restriction of coarse-grained measurement. However, while considering the violation of Bell-CHSH inequality, they have studied the effect of either coarse-graining of final measurement resolution or the coarsening of measurement references. Thus, the question remained open that, if one considers complete coarse-grained measurement, characterized by consideration of both aspects of coarse-graining simultaneously, then to what extent the \textit{quantumness} of the system can persist? Furthermore, one can ask whether there exists a way through which one can witness the \textit{quantumness} of a macroscopic entangled state, with a coarse-grained measurement that is characterized by a larger degree of imprecision than that depicted in \cite{H.Jeong}. In this paper, while we find the answer to the former question, we assert that the latter can be answered affirmatively.

In this work, we consider a bipartite scenario with two spatially separated parties, where each of them shares one part of a nontrivially correlated system. Thus, in this case, the relevant nonclassical correlations namely Bell nonlocality and quantum steering are captured through the violation of suitable Bell inequalities and linear steering inequalities respectively. We investigate the effect of coarsening of final measurement resolution as well as the measurement reference, on the quantum violation of a family of symmetric Bell inequalities \cite{N.Gisin1999} with an increasing number of measurement settings on each side. For the odd number of settings, it is seen that the effect of coarse-graining on quantum-to-classical transition can be compensated by testing a Bell inequality with an increased number of settings. However, for the even number of settings, we witness an opposite trend. Due to this lack of consistency of nonlocal correlations captured through the symmetric Bell inequalities in revealing the quantum-to-classical transition, we next consider steering as the viable quantum correlation. We explicitly show that when steering is taken as the relevant nonclassical feature of a specific macroscopic entangled state, with respect to a steering inequality having a particular number of measurement settings, if the quantum-to-classical transition occurs at a specific degree of coarsening, then by increasing the number of settings with the new steering inequality, one can still witness the quantum violation for the same degree of coarsening. Thus, unlike nonlocality in the case of steering for any arbitrary number of measurement settings, the effect of coarsening of measurement can be narrowed down by increasing the number of measurement settings.

This paper is organized as follows. In Sec.II, we have discussed the operational description of different forms of bipartite quantum correlation briefly and the coarse-grained measurement elaborately. In Sec.III, we study the quantum-to-classical transition of macroscopic entangled states under complete coarse-grained measurement by considering Bell nonlocality as a defining notion of classicality. Next, in Sec.IV, we move on to another nonclassicality witness namely the quantum steering and study the quantum-to-classical in terms of coarsening of measurement. Then in Sec.V, we extend our study for mixed macroscopic entangled state. Finally we conclude the paper in Sec.VI by discussing the implications of the current work and future directions.

\section{Prerequisites}
\subsection{Layers of non-classicality in bipartite scenario}
Before going into the discussion of the effect of coarsening of measurements on the \textit{quantumness} exhibited by a macroscopic entangled state let us first describe different forms of bipartite quantum correlations relevant to this paper. In order to discuss these different forms of quantum correlations namely the Bell nonlocality, quantum steering, and entanglement, let us consider   $\mathcal{M}^A$ and $\mathcal{M}^B$ as the set of all observables, in the Hilbert spaces of the subsystems that are in possession of of two spatially separated parties namely Alice’s and Bob respectively. The operators $\mathcal{A}_{x} \subseteq \mathcal{M}^{A}$ and $\mathcal{B}_{y} \subseteq \mathcal{M}_{B}$, represent the collection of measurement operators, corresponding to the input measurement settings $x$ and $y$ of Alice and Bob, with respective outcomes denoted by the variables $a \in \mathbb{R}_{A}$ and $b \in \mathbb{R}_{B}$. The set of ordered pairs, $\mathcal{S} = \{ (\mathcal{A}_{x},\mathcal{B}_{y}): \mathcal{A}_{x} \in \mathcal{M}^{A}, \mathcal{B}_{y} \in \mathcal{M}^{B} \}$, denote a \textit{measurement strategy}. A behavior is defined as joint probabilities for all outcomes of all pairs of measurement observables in a measurement strategy for a given state. Alice and Bob perform measurements on their respective subsystem originated from spatially separated non-trivially correlated systems denoted by the density matrix $\rho$ such that  the joint probability of obtaining outcomes $a$ and $b$ upon measuring $\mathcal{A}_{x}$ and $\mathcal{B}_{y}$ is given as,
\begin{equation}\label{eqprb}
    P(a,b|x,y,\rho) = \text{Tr}[\rho(\mathcal{A}_{x} \otimes \mathcal{B}_{y})].
\end{equation}

\textit{\textbf{Bell nonlocality}}: The strongest form of bipartite quantum correlation exhibited by nontrivially correlated and specially separated systems is nonlocality. A behavior is said to have a local-realistic
model, if and only if for all $\mathcal{A}_{x} \in \mathcal{M}_{A}$, $a \in \mathbb{R}_{A}$, $\mathcal{B}_{y} \in \mathcal{M}_{B}$, $b \in \mathbb{R}_{B}$, there exist probability distributions $P(\lambda)$,  $P(a|x,\lambda)$ and $P(b|y,\lambda)$, involving the local hidden variable $\lambda$ for both Alice and Bob that reproduce the joint probabilities of Eq.(\ref{eqprb}) in the form,
\begin{equation}\label{eqbn}
P(a,b|x,y,\rho) = \sum_{\lambda} P(\lambda) P(a|x,\lambda) P(b|y,\lambda),
\end{equation}

where, $P(a|x,\lambda)$ represents probability distributions of outcomes $a$ on the measurement of $\mathcal{A}_{x}$ which can be reproduced from a local hidden variable $\lambda$. The probability of obtaining an outcome $a$ on the measurement of $\mathcal{A}_{a|x}$ is freely determined by the hidden variables $\lambda$ and similar definitions apply to Bob's systems. Using the set of all possible behaviors any constraint that can be obtained from Eq.(\ref{eqbn}) is called a Bell inequality. A state for which all the behavior can be given a local-realistic description is called a Bell local state. Any state that doesn't exhibit the characteristics of a Bell local state is called Bell nonlocal state \cite{N.Brunner,H.M.Wiseman,S.J.Jones,E.G.Cavalcanti}.

\textit{\textbf{Quantum steering}:} A behavior is described by a local hidden state (LHS) model, if and only if for all $\mathcal{A}_{x} \in \mathcal{M}_{A}$, $a \in \mathbb{R}_{A}$, $\mathcal{B}_{y} \in \mathcal{M}_{B}$, $b \in \mathbb{R}_{B}$, there exist probability distributions  $P(\lambda)$,  $P(a|x,\lambda)$ and $P_{Q}(b|y,\lambda)$, characterised by variable $\lambda$ for Alice and local hidden state $\rho_{B}(\lambda)$ for Bob that reproduce the joint probabilities of Eq.(1) in the form,
\begin{equation}\label{eqstr}
P(a,b|x,y,\rho) = \sum_{\lambda} P(\lambda) P(a|x,\lambda) P_{Q}(b|y,\lambda)
\end{equation}

where, $P_{Q}(b|x,\lambda)$ represents probability distributions of Bob's outcomes $b$ on the measurement of $\mathcal{B}_{y}$ which are compatible with a local hidden quantum state $\rho_{B}(\lambda)$. In quantum theory the probability of obtaining an outcome $b$ on the measurement of  $\mathcal{B}_{b|y}$, is given by $P_{Q}(b|y,\lambda) = \text{Tr}[\rho_{B}(\lambda)\mathcal{B}_{b|y}]$. However, Alice’s outcomes are free to be arbitrarily determined by the variables $\lambda$. Exhausting the set of all possible behaviors any form of constraint that can be derived from Eq.(\ref{eqstr}) is called a quantum steering inequality. A state for which all the behaviors can be given a LHS model is called a quantum unsteerable state. Any state that doesn't exhibit the properties of LHS model is called a steerable state \cite{R.Uola,H.M.Wiseman,S.J.Jones,E.G.Cavalcanti} and this particular phenomena is called quantum steering.

\textit{\textbf{Entanglement}}: A behavior is said to describe a separable system, if and only if for all $\mathcal{A}_{x} \in \mathcal{M}_{A}$, $a \in \mathbb{R}_{A}$, $\mathcal{B}_{y} \in \mathcal{M}_{B}$, $b \in \mathbb{R}_{B}$, there exist probability distributions $P(\lambda)$, $P_{Q}(a|x,\lambda)$ and $P_{Q}(b|y,\lambda)$, involving the variable $\lambda$ and Alice's and Bob's local hidden quantum states $\rho_{A}(\lambda)$ and $\rho_{B}(\lambda)$ respectively that reproduce the joint probabilities of Eq.(\ref{eqprb}) in the form,
\begin{equation}\label{eqent}
    P(a,b|x,y,\rho) = \sum_{\lambda} P(\lambda) P_{Q}(a|x,\lambda) P_{Q}(b|y,\lambda)
\end{equation}
where, $P_{Q}(a|x,\lambda)$ and $P_{Q}(a|x,\lambda)$ represents the probability distributions of outcomes $a$ and $b$ on the measurement of $\mathcal{A}_{x}$ and $\mathcal{B}_{y}$ which are compatible with local hidden quantum states described by  $\rho_{A}(\lambda)$ and $\rho_{B}(\lambda)$. The probability of obtaining an outcome $a$ on the measurement of $\mathcal{A}_{a|x}$, is given by $P_{Q}(a|x,\lambda) = \text{Tr}[\rho_{A}(\lambda)\mathcal{A}_{a|x}]$ and similar definitions apply for Bob's subsystems.\\

Any form of constraint on the set of all possible behavior that can be derived from Eq.(\ref{eqent}) is called a separability criterion. A state for which all the behaviors can be given a LHS-LHS model is called a separable state. Any state is called entangled or nonseparable if it doesn't exhibit the characteristics imposed by the separability criterion \cite{R.Horrodecki,H.M.Wiseman,S.J.Jones,E.G.Cavalcanti}.

Among these aforementioned correlations Bell nonlocality is the strongest form of bipartite quantum correlation and entanglement is the weakest form of quantum nonlocal correlation characterised by LHS-LHS model, i.e., entanglement can be described by weakening the description of Bell nonlocality through a local hidden state model for both Alice and Bob respectively. Therefore, Bell nonlocal states are a subset of entangled states \cite{H.M.Wiseman}. Quantum steering is a weaker form of nonlocality lying intermediate between Bell nonlocality and entanglement that is characterized by LHS model, i.e., quantum steering can be described by weakening the description of Bell nonlocality via a local hidden state model for Bob or by fine graining the description of entanglement via a local hidden variable for Alice. Thus Bell nonlocal states are a subset of steerable states, and steerable states are subset of entangled states \cite{H.M.Wiseman}.

An important point to note is that the pure entangled states display no distinction among the different forms of bipartite quantum correlations \cite{R.F.Werner,N.Gisin1991,S.Popeschu}. However, this is not the case for mixed states, i.e., not all states that are entangled display steering \cite{H.M.Wiseman}, and not all states that are steerable displays Bell nonlocality \cite{R.F.Werner}. For example, two-qubit Werner states which is given as, $\rho_W = p|\phi^{-} \rangle\langle\phi^{-}| + \frac{1-p}{4} \mathbb{1}$, where $|\phi^{-}\rangle = \frac{1}{\sqrt{2}}(|01\rangle-|10\rangle)$ and $\mathbb{1}$ is the identity matrix. $\rho_{W}$ is Bell nonlocal for $p > \frac{1}{\sqrt{2}}$ \cite{M.Horodecki, A.Peres}, steerable for $p > \frac{1}{2}$ \cite{H.M.Wiseman} and entangled for $p > \frac{1}{3}$ 
 \cite{R.F.Werner}.
\subsection{Preliminaries on coarse-grained measurements}
A complete measurement process comprises of two different parts; the first part involves an interaction that entangles the system with the measurement apparatus, we shall call this process as setting the measurement reference. Depending on the observable one intend to measure, this process in quantum theory is usually represented by a unitary operation which prepare the system in such a way that each system state becomes correlated with each of the pointer state in a unique way. On the other hand, the second part involves inferring the value of the observable by observing the pointer state of the measurement apparatus. In quantum theory, such detection corresponds to the eigenvalue of the measurement operator representing the concerned observable. However, due to lack of efficiency of the measurement apparatus, the actual situation usually remains far from such idealizations and we can only extract information about the system up to a certain degree of coarse-grained version of both the entangling interaction as well as the final detection. This is explicitly discussed in the following.

In order to contemplate the physical grounds behind the coarse-grained measurement let us first consider a multilevel quantum system represented in the following way,

\begin{equation}
    \ket{\Psi_{in}}= \sum_{\text{n}}l_{n}\ket{l_{n}},
\end{equation}

where $\{\ket{l_{n}}\}$ corresponds to orthonormal eigen basis of some operator $\hat{F}$. Besides the state of the system, we consider the initial state of the measuring pointer to be a Gaussian wave packet represented in the position basis as,

\begin{eqnarray}
\label{eq:pnti}
    \ket{\Phi_{pointer}^{i}} &&= \sum_{q} \phi(q)\ket{q} \nonumber \\
    &&=  \frac{1}{\sigma \sqrt{2 \pi}} \sum_{q} exp \big(-\frac{q^2}{2\sigma^{{2}}}\big) \ \ket{q}.
\end{eqnarray}
This Gaussian distribution has a peak (mean) at the $q=0$ and a standard deviation $\sigma$. Thus the combined state of the beam and the pointer is, $\ket{\Psi} = \ket{\Psi_{in}} \otimes \ket{\Phi_{pointer}^{i}}$. The interaction between the system and measuring pointer is mathematically represented as a unitary, $\mathbb{\hat{U}}(\epsilon)= e^{-i\epsilon \hat{H} }$, with the effective Hamiltonian as, $\hat{H}= \hat{F} \otimes  \hat{p}$. Here $\hat{F}$ is the operator corresponding to the dynamical variable we want to measure, $\epsilon$ represents the strength of measurement interaction, and $\hat{p}$ is the operator corresponding to the variable conjugate to $q$.

After the system interacts with the measuring apparatus the state of the combined system becomes,

\begin{eqnarray}
    \mathbb{\hat{U}}(\epsilon) \ket{\Psi} &&=  e^{-i\epsilon (\hat{F} \otimes  \hat{p})  } \ket{\Psi} \nonumber \\
    &&= e^{-i\epsilon (\hat{F} \otimes  \hat{p})  } \big( \sum_{n}l_{n}\ket{l_{n}} \big) \otimes \big( \sum_{x} \phi(q)\ket{q} \big) \nonumber \\
    &&=  \frac{1}{\sigma \sqrt{2 \pi}} \sum_{n}l_{n}\ket{l_{n}} \otimes \sum_{x} exp \big(-\frac{(q-\epsilon l_{\text{n}})^2}{2\sigma^{{2}}}\big) \ \ket{q}. \nonumber \\
\end{eqnarray}
Thus, after the interaction, the initial pointer state's wave function is shifted by different amounts depending on the eigenvalues $l_{n}$. Moreover, as long as the the interaction strength $\epsilon$ is substantially large and the standard deviation $\sigma$ remains less than the separation between the eigenvalues, the peaks of the Gaussian appear to be well resolved from one another. However, as the standard deviation becomes large compared to the consecutive eigenvalue ranges, the shifted Gaussian profiles due to different eigenvalues overlap making it difficult to identify individual shifts. In such situations, only a group of peaks can be resolved from another group leading to the coarsening of the measurement resolution as described below.

\textit{Coarsening of measurement resolution:} Consider a dichotomic observable in arbitrary dimension  as $\hat{F}_{k} = \hat{F}_{k}^{+} - \hat{F}_{k}^{-}$ such that
\begin{align}\label{eq1}
    \hat{F}_{k}^{+} &= \sum_{n = k+1}^{\infty}|l_{n}\rangle \langle l_{n}| & \hat{F}_{k}^{-} &= \sum_{n =-\infty}^{k}|l_{n}\rangle \langle l_{n}|,
\end{align}

where $\hat{F}_{k}^{\pm}$'s are projectors of $\hat{F}_{k}$ corresponding to the eigenvalues $\pm1$. It is important to note that, due to limited resolution individual eigenvectors are clubbed into two projectors with a sharp boundary at $k$. An unsharp (fuzzy) version of the same measurement operator can be written as,
\begin{equation}\label{eq2}
    \hat{F}_{\delta} = \sum_{k = -\infty}^{\infty}P_{\delta}(k)\hat{F}_{k},
\end{equation}

where $P_{\delta}(k) = \sum_{-\infty}^{\infty}\frac{1}{\delta\sqrt{2\pi}}e^{\frac{-k^{2}}{2\delta^{2}}}$ is the normalized discrete Gaussian kernel. Since we use a discrete Gaussian distribution function, we assume $\delta > 1$ for the normalization conditions. Here $\delta$ represents the degree of coarsening of the final measurement resolution. Any two states $|l_{n}\rangle$ and $|l_{-n}\rangle$ of a high Hilbert space dimension are called macroscopically distinguishable under such fuzzy measurement if they can be discriminated with a high success probability given by,
\begin{equation}
    P_{s} = \left|\sum_{k=-\infty}^{\infty}\frac{1}{\delta\sqrt{2\pi}}e^{\frac{-k^{2}}{2\delta^{2}}}\zeta_{n-k}\right|^{2},
\end{equation}
where $\zeta_{n-k} = 1$ for $n-k > 0 $ and $-1$ for $n-k \leq 0$.

An entangled state prepared with such microscopically distinguishable states can be represented as follows,
\begin{equation}\label{eq3}
    |\Psi_{n}\rangle = \frac{|l_{n}\rangle|l_{-n}\rangle + |l_{-n}\rangle|l_{n}\rangle}{\sqrt{2}}.
\end{equation}

This is called a macroscopic entangled state when n is sufficiently large. The term 'macroscopic' was first introduced in \cite{H.Jeong} has also been used in recent literature \cite{M.Chaudhary}, denotes a system with high dimensionality rather than a low-dimensional system with a large mass or size.

As already discussed, the value of an observable that we want to measure corresponds to the detection of some entity on the measuring apparatus entangled with the system. This detected entity essentially lies in the real configuration space specified through the measurement apparatus itself. In ideal sharp measurement, the one-to-one correspondence between the eigenstates of the observable and the detected value of the entity corresponding measurement outcome enables one to extract complete information about the system by maximally disturbing it. In contrast, under coarsening of the final resolution of measurements this correspondence is lost i.e., a given detection in the measurement apparatus can correspond to more than one eigenstate of the concerned observable. Hence, the coarse-grained measurement fails to perfectly detect the eigenvalues of the observable. However, in such a measurement, the disturbance to the system is also less compared to the ideal sharp measurement. Thus any remaining quantum correlation after such measurement becomes a function of the degree of coarsening of that particular measurement. This motivates one to study the nature of different quantum correlations under the coarsening of measurement resolution.

\textit{Coarsening of measurement reference:} 
In a particular measurement process, depending on what observable one wishes to measure the unitary evolution characterised by the interaction Hamiltonian essentially rotates the system state in the direction of a particular reference vector on the Bloch sphere. This reference and thereby the rotation is different for different measurements. For example let us consider the unitary $\hat{\mathcal{U}}(\phi)$ whose action is to rotate two orthonormal basis states $|l_{n}\rangle$ and $|l_{-n}\rangle$ according to,
\begin{align}\label{eq4}
    \hat{\mathcal{U}}(\phi)|l_{n}\rangle &= cos\phi|l_{n}\rangle + sin\phi|l_{-n}\rangle, \\ \nonumber
    \hat{\mathcal{U}}(\phi)|l_{-n}\rangle &= sin\phi|l_{n}\rangle - cos\phi|l_{-n}\rangle,
\end{align}
where $\phi$ is unique for a particular measurement. This $\phi$ is a function of the measurement operator as well as the strength of interaction discussed in the previous subsections. A coarse-grained version of the unitary operation applied to the measurement operators $\hat{F}_{\delta}$ can be described as follows,
\begin{equation}\label{eq:ref}
    \hat{F}_{\delta,\Delta}(\phi_{o}) = \int d\phi P_{\Delta}(\phi-\phi_{o})\left(\hat{U}(\phi)\hat{F}_{\delta}\hat{U}^{\dag}(\phi)\right)
\end{equation}

where, $P_{\Delta}(\phi-\phi_{o})= \frac{1}{\Delta \sqrt{2\pi}}e^{\frac{-(\phi_{i}- \theta_{i})^{2}}{2\Delta^{2}}}$ is the Gaussian Kernel centred around $\theta_{i}$ with a normal deviation $\Delta$. Here, $\Delta$ is a measure of the degree of coarsening in the measurement reference.

The operation of setting the input or the reference of the measurement apparatus may not always be precise. Therefore, an inaccurate control of the reference leads to a setting of the measurement apparatus other than the desired one. Only under a specific choice of measurement operators (incompatible) and a specific amount of imprecision, the nonclassical correlations are observed in terms of certain restrictions imposed through some nonclassicality witness. However, a random choice of any of the two would not lead to the observation of nonclassical correlations. We study the effect of coarsening of both measurement references as well as the final resolution of the measurement in the quantum-to-classical transition of the macroscopic entangled states in terms of Bell nonlocality and quantum steering.

\section{Effect of complete coarse-grained measurement on nonlocal correlations}
The quantum-to-classical transition has been studied \cite{H.Jeong} in the context of coarse-grained measurement defining local-realism as classicality and exploring the quantum violation of Bell-CHSH inequality \cite{J.F.Clauser} by a macroscopic entangled state. It was demonstrated in \cite{H.Jeong} that the point of quantum-to-classical transition corresponding to the degree of coarsening of final measurement resolution is dependent on the macroscopicity $n$ of the entangled state. For example, the degree of coarsening for which a system represented by a macroscopic entangled state with  $n =2$ satisfies the Bell-CHSH inequality, the same amount of coarsening may offer a quantum violation of the inequality with states having a higher value of $n$. In other words, the quantum mechanical violation of the Bell-CHSH inequality persists with increasing degrees of fuzziness as $n$ increases. In contrast, when measurement reference is coarsened, quantum-to-classical transitions occur regardless of the degree of macroscopicity $n$, implied by the fact that the amount of Bell-CHSH violation in this case is independent of the value of n. Importantly, the Bell-CHSH violation depends on the incompatibility \cite{brub23,wolf09,bene18,hirsch18} between the local measurement observables, therefore, to find robustness of the nonlocality under complete coarse-grained measurement we consider nonlocality witnesses that incorporate an arbitrary number of measurements instead of only two measurements per side as in the case of standard Bell-CHSH inequality.

In order to investigate the tensility of the \textit{quantumness} exhibited by the aforementioned macroscopic entangled states under a complete coarse-grained measurement, let us introduce the following family of symmetric Bell inequalities \cite{N.Gisin1999} with an arbitrary number of measurement settings,
\begin{multline}\label{eq1d}
    \mathcal{B}_{m} \equiv \sum_{i=1}^{m}\left(\sum_{j=1}^{m+1-j} \langle\hat A_{i},\hat B_{j} \rangle-\sum_{j=m+2-i}^{m} \langle \hat A_{i},\hat B_{j} \rangle \right) \\ \le \left[\frac{{m}^{2}+1}{2}\right] \equiv \mathcal{L}_{m}
\end{multline}

where we call $\mathcal{B}_{m}$ as the nonlocality witness with m measurements per party in a bipartite scenario and $[r]$ denotes the largest integer smaller or equal to $r$. $\hat{A}_{i}$ with $i\in \{1,m\}$ and $\hat{B}_{j}$ with $j\in \{1,m\}$ are Alice's and Bob's observables respectively, and $\mathcal{L}_{m}$ is the bound on the nonlocality witness $\mathcal{B}_{m}$ imposed by local realism. For $m=2$ settings, Eq.\eqref{eq1d} reduces to standard Bell-CHSH inequality. Since the violation of inequality given by Eq.\eqref{eq1d} reveals the intrinsic nonlocality of the macroscopic entangled state under fuzzy measurements, the violation tends to decrease and at a certain degree of fuzziness of measurement, the statistics can no longer violate the inequality. We refer to this as the quantum-to-classical transition of the macroscopic entangled state. As discussed in the earlier section the fuzziness in measurement can come either from coarse-graining of the final resolution of measurement outcomes or due to coarsening of measurement reference or the combined effect of both of them. In the following we explicitly showcase the the effect of each of these differently originated fuzzy measurements in quantum-to-classical transition.

(\textbf{\rom{1}}) First,  we consider the effect of fuzziness in the final resolution of the outcome of measurement characterized by $\delta > 1$, with perfect control on the measurement reference ($\Delta=0$). In such case the fuzzy version of the bipartite correlation function,  $\langle\hat{A}_{i} \hat{B}_{j}\rangle_{\delta}= \langle\hat{A}_{\delta} (\phi_{i}) \otimes  \hat{B}_{\delta} (\phi_j)\rangle$, due to the measurements $A_{i}$ and $B_{j}$ performed respectively by Alice and Bob on their subsystem from the shared system, represented by $\ket{\Psi_{n}}$ can be written in a simplified form as follows,
\begin{eqnarray}\label{eq9}
   \langle\hat{A}_{i} \hat{B}_{j}\rangle_{\delta}
     &=& \frac{1}{2}[ \mathcal{Q}_{\delta}(n,\phi_{i})\hspace{0.1cm}\mathcal{Q}_{\delta}(-n,\phi_{j}) + \mathcal{Q}_{\delta}(-n,\phi_{i})\hspace{0.1cm}\mathcal{Q}_{\delta}(n,\phi_{j}) \nonumber \\
&& \hspace{2.7cm} +  2\mathcal{R}_{\delta}(n,\phi_{a})\hspace{0.1cm}\mathcal{R}_{\delta}(n,\phi_{b}) ], 
\end{eqnarray}
where $\phi_{i}$ and $\phi_{j}$ denotes the angle with z-axis of a Bloch sphere representing the direction of measurements for the observable $A_{i}$ and $B_{j}$ respectively, and $\mathcal{Q}_{\delta}(n,\phi) = \sum_{k=-\infty}^{\infty} P_{\delta}(k)\hspace{0.1cm}[cos^{2}\phi\hspace{0.1cm}\zeta_{n-k}+sin^{2}\phi\hspace{0.1cm}\zeta_{-n-k})]$ and $\mathcal{R}_{\delta}(n,\phi) = sin\phi \hspace{0.1cm} cos\phi \sum_{k=-\infty}^{\infty} P_{\delta}(k)\hspace{0.1cm}[\zeta_{n-k}-\zeta_{-n-k}]$.

\begin{figure}[H]
  \centering
  \includegraphics[scale=0.46]{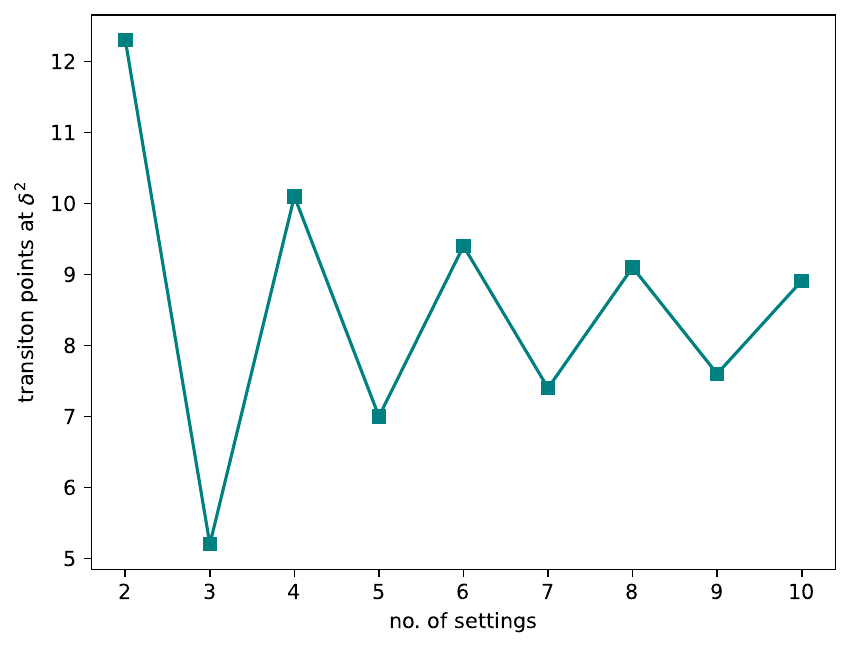}
\caption{ In the figure the Value of $\delta^{2}$ corresponding to the transition points (for Bell nonlocality) is plotted against the numbers of measurement settings with the degree of macroscopicity $n=5$ of the entangled state.}
  \label{inconsistency}  
\end{figure}

\begin{figure}[h]
\centering
\subfloat[]{\includegraphics[width = 0.25 \textwidth, height=0.18\textwidth]{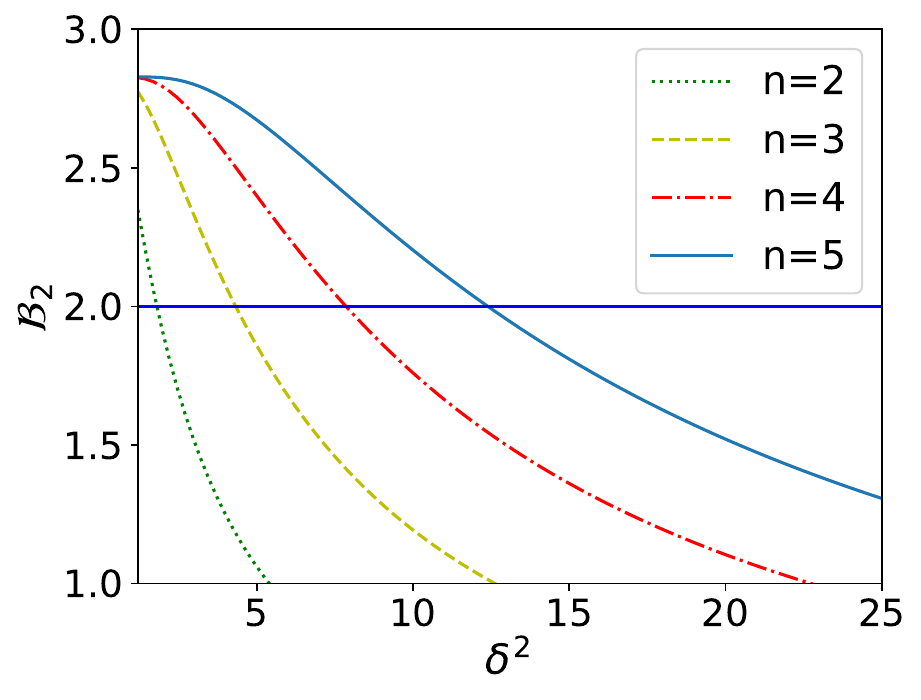}} 
\subfloat[]{\includegraphics[width = 0.25 \textwidth, height=0.18\textwidth]{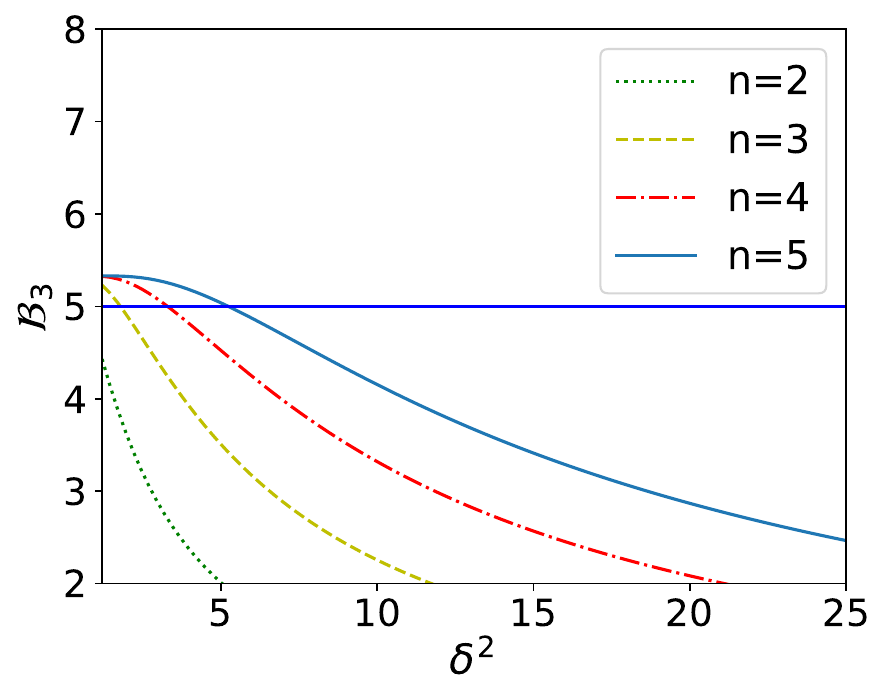 }} \\
\subfloat[]{\includegraphics[width = 0.25 \textwidth, height=0.18\textwidth]{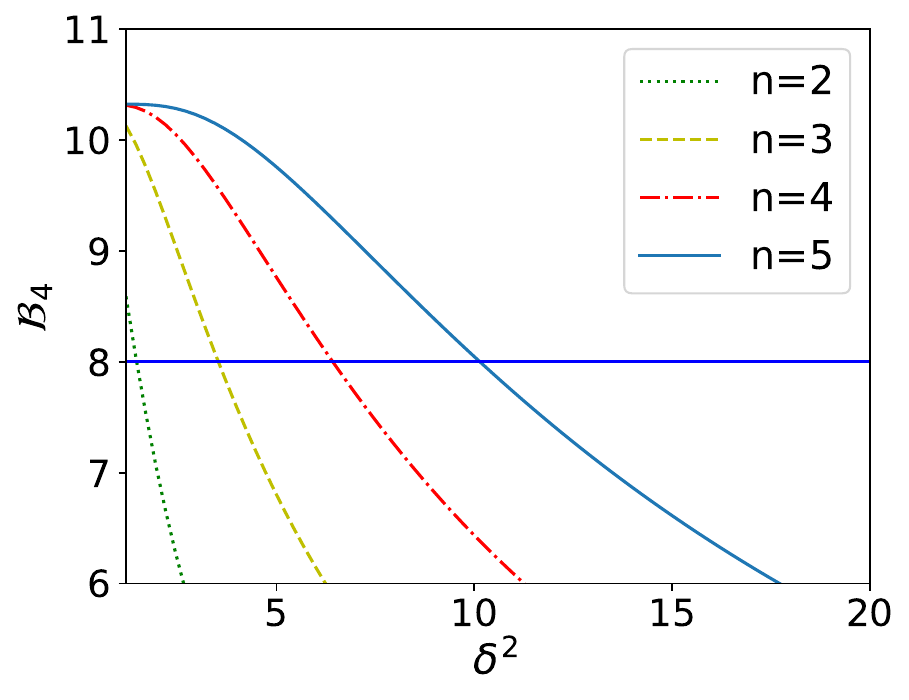 }}
\subfloat[]{\includegraphics[width = 0.25 \textwidth, height=0.18\textwidth]{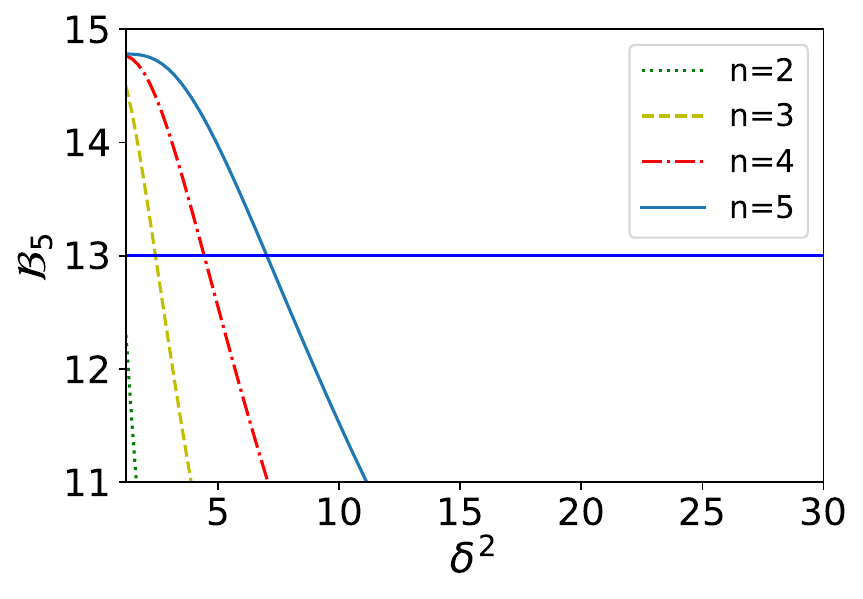 }}
\caption{In this figure, each plot represents numerically optimized nonlocality witness $\mathcal{B}_m$ has been plotted against the variance $\delta^{2}$ for different value of $n$. The blue line represents the classical limit in each case. Input settings $m$ = 2, 3, 4, and 5 are plotted in (a),(b), (c), and (d) respectively. }
\label{fig_bl}
\end{figure}

In order to introduce the effect of coarsening on nonlocal correlations we replace the expectation values appearing in Eq.\eqref{eq1d} by its fuzzy counterpart and numerically optimized nonlocality witness $\mathcal{B}_{m}$ for different values of the degree of macroscopicity $n$. For a particular macroscopic entangled state (with fixed $n$) the value of nonlocality witness $\mathcal{B}_{m}$ depends on the degree of coarse-graining $\delta$. Moreover, if $\delta_{n,m}^{c}$ denotes the degree of coarsening for which $\mathcal{B}_{m}$ stays above the classical local-realist bound for the state $\ket{\Psi_n}$, then for certain value of $\delta$ the state no longer exhibit non-classical behavior only if $\delta \leq \delta_{n,m}^{c}$. The value $\delta_{n,m}^{c}$ for which the system stops behaving quantum mechanically and starts behaving classically will be called the transition point. We observe that for the nonlocality witnesses having an even number of measurement settings $(m=2,4,6...)$, with an increase in the number of settings $m$, the system demonstrates the quantum-to-classical transition at a lower value of measurement fuzziness ($\delta$). More precisely, for a fixed $n$, the transition points tend to decrease as the number of settings increases, viz.,  $\delta_{n,m=2}^{c} > \delta_{n,m=4}^{c} > \delta_{n,m=6}^{c}$, etc. In contrast, for the nonlocality witness having odd number of measurement settings $(m=3,5,7..)$, with an increase in $m$ the transition points follows an opposite trend as  $\delta_{n,m=3}^{c} < \delta_{n,m=5}^{c} < \delta_{n,m=7}^{c}$, etc. This trend is explicitly depicted in Fig. \ref{inconsistency}.
In the Fig.\ref{fig_bl}, we have plotted the value of $\mathcal{B}_{m}$ as a function of $\delta^{2}$ for macroscopic entangled states characterized by different values of $n$. We observe that for $m=2$ the nonlocality witness stays above the local-realist bound for different ranges of $\delta$ depending on the value of $n$. This was already noted in ref. \cite{H.Jeong}. However, the question arises how robust is the \textit{quantumness} of a specific macroscopic entangled state (with fixed value $n$) under different nonlocality witnesses characterized by an arbitrary number of measurement settings $m$. We found that for a certain value of the degree of coarsening $\delta$ the value of $\mathcal{B}_{m}$ decreases as the number of setting $m$ increases when $m$ is even. On the other hand, for odd $m$ we see the opposite trend, i.e, the value of $\mathcal{B}_{m}$ increases for all even $m$ as the degree of coarsening increases. Thus, for any odd number of settings, the nonlocality witness significantly captures the tensility of the critical point $\delta_{n,m}^{c}$ for the quantum-to-classical transition of the macroscopic entangled state under coarsening of final measurement resolution. However, any such witness with an even number of settings fails to capture the essence of the incremental nature of the \textit{quantumness} with the increase in the number of settings.
\\
(\textbf{\rom{2})} Now, we consider the situation where the final outcomes of the measurement are perfectly resolved ($\delta=0$) but the measurement reference represented by the unitary transformation is coarsened with degree of coarsening $\Delta$. So, the joint correlation function is now dependent on $\Delta$ and can be written in the following form,
\begin{multline}\label{eq10}
     \langle\hat{A}_{i} \hat{B}_{j}\rangle_{\Delta}= -\int_{-\infty}^{\infty}\int_{-\infty}^{\infty}d\phi_{i}d\phi_{j}P_{\Delta}(\phi_{i}-\theta_{i})\\ \times P_{\Delta}(\phi_{j}-\theta_{j})\cos{[2(\phi_{i}+\phi_{j})]}
\end{multline}

where, $P_{\Delta}(\phi_{j}-\theta_{j})= \frac{1}{\Delta \sqrt{2\pi}}e^{\frac{-(\phi_{i}- \theta_{i})^{2}}{2\Delta^{2}}}$ is the Gaussian Kernel centred around $\theta_{i}$ with a normal deviation $\Delta$. Unlike the previous case, the correlations in the LHS of Eq.(\ref{eq10}) is not a function of $n$, suggesting that the quantum-to-classical transition of the macroscopic entangled state resulting from coarsening of measurement reference is independent of the variable $n$. Thus no matter how large macroscopicity the system possesses, the system defies any nonlocal behavior as soon as the degree of coarsening of measurement reference reaches a specific value $\Delta_{n,m}^{c}$. However, this is not the whole story, as we will see in the following that the number of measurement settings used in the nonlocality test plays a significant role in shifting the value $\Delta_{n,m}^{c}$.

(\textbf{\rom{3}}) Finally, we consider what we call the complete coarse-grained measurement in which the effect of fuzziness in both the final measurement resolution as well as the measurement reference is considered. Now, under the complete coarse-grained measurement, the fuzzy version of the correlation functions can be written as,
\begin{equation}\label{Ecor}
    \langle A_{i},B_{j} \rangle_{\delta, \Delta} = \left\langle \hat{A}_{\delta,\Delta}(\phi_{i})\otimes\hat{B}_{\delta,\Delta}(\phi_{j})\right\rangle,
\end{equation}

where $\phi_{i}$ and $\phi_{j}$ as mentioned earlier are the angles with the z-axis of a Bloch sphere representing the direction of measurements for the observable $A_{i}$ and $B_{j}$ respectively. Substituting $\hat{A}_{\delta,\Delta}(\phi_{i})$ and $\hat{B}_{\delta,\Delta}(\phi_{j})$ from Eq.\eqref{eq:ref} we can write Eq. \eqref{Ecor} as,
\begin{eqnarray}\label{eqccg}
    \hspace{-0.5cm} \langle A_{i},B_{j} \rangle_{\delta, \Delta} & = & \int_{-\infty}^{\infty}\int_{-\infty}^{\infty}d\phi_{a}d\phi_{b}P_{\Delta}(\phi_{a}-\theta_{a})P_{\Delta}(\phi_{b}-\theta_{b}) \nonumber \\
&&\left\langle(U^\dagger(\phi_{a})F_{\delta}U(\phi_{a}))\otimes(U^\dagger(\phi_{b})F_{\delta}U(\phi_{b}))\right \rangle.
\end{eqnarray}

Finally, the fuzzy version of the correlation function can be written in a more simplified form as,

\begin{equation} \label{eq:corfinal}
 \langle A_{i},B_{j} \rangle_{\delta, \Delta} =  \int_{-\infty}^{\infty}\int_{-\infty}^{\infty}d\phi_{1}d\phi_{2}P_{\Delta}(\phi_{1}-\theta_{1})P_{\Delta}(\phi_{2}-\theta_{2})\langle\hat{A}_{i} \hat{B}_{j}\rangle_{\delta},
\end{equation}

where, $\langle A_{i},B_{j} \rangle_{\delta, \Delta}=\frac{1}{2}[ \mathcal{Q}_{\delta}(n,\phi_{i})\hspace{0.1cm}\mathcal{Q}_{\delta}(-n,\phi_{j}) + \mathcal{Q}_{\delta}(-n,\phi_{i})\hspace{0.1cm}\mathcal{Q}_{\delta}(n,\phi_{j}) +  2\mathcal{R}_{\delta}(n,\phi_{a})\hspace{0.1cm}\mathcal{R}_{\delta}(n,\phi_{b}) ].$

\begin{figure}[h]
\centering
\subfloat[]{\includegraphics[width = 0.25 \textwidth, height=0.19\textwidth]{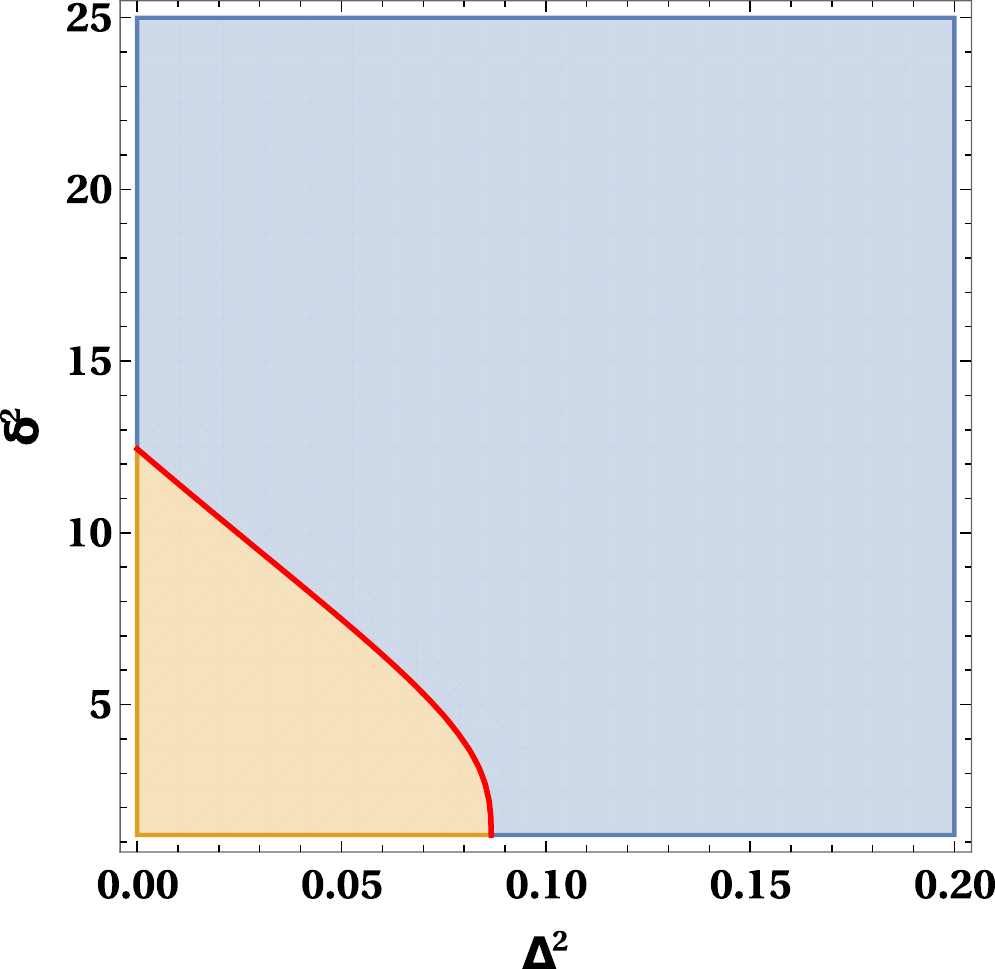}} 
\subfloat[]{\includegraphics[width = 0.25 \textwidth, height=0.19\textwidth]{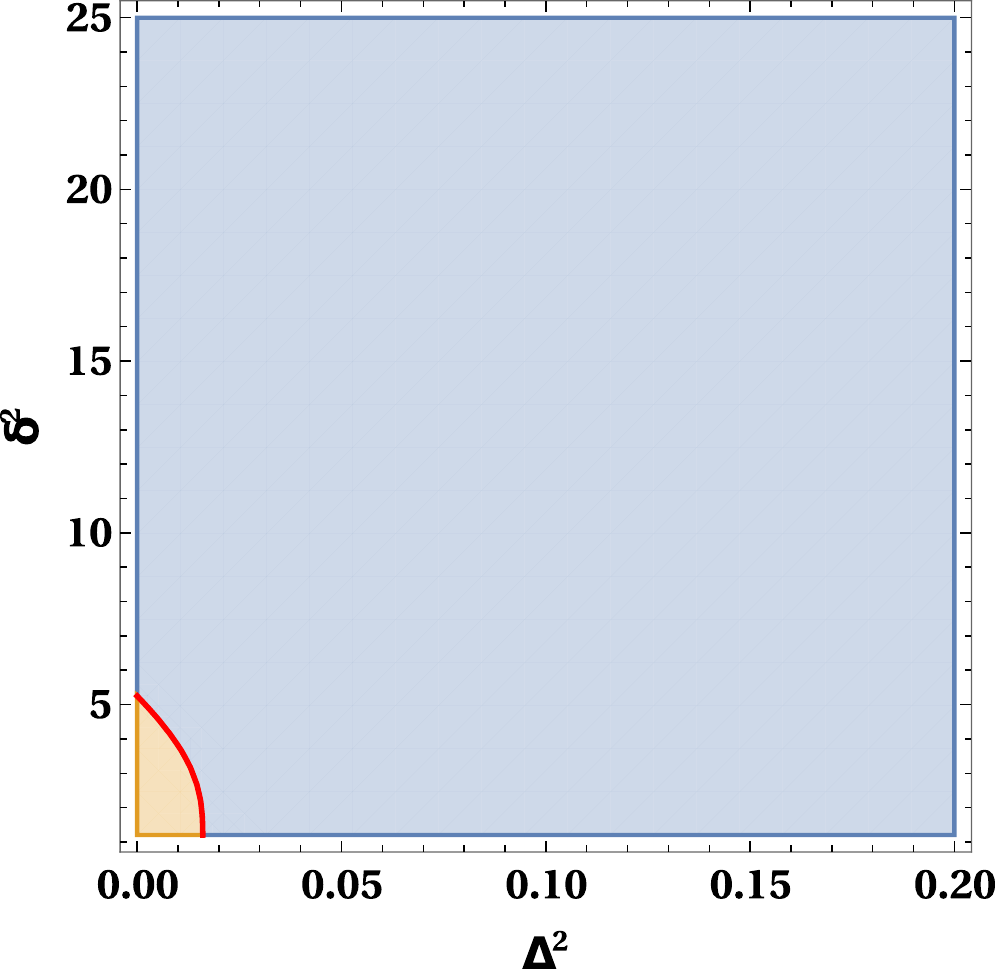}} \\
\subfloat[]{\includegraphics[width = 0.25 \textwidth, height=0.19\textwidth]{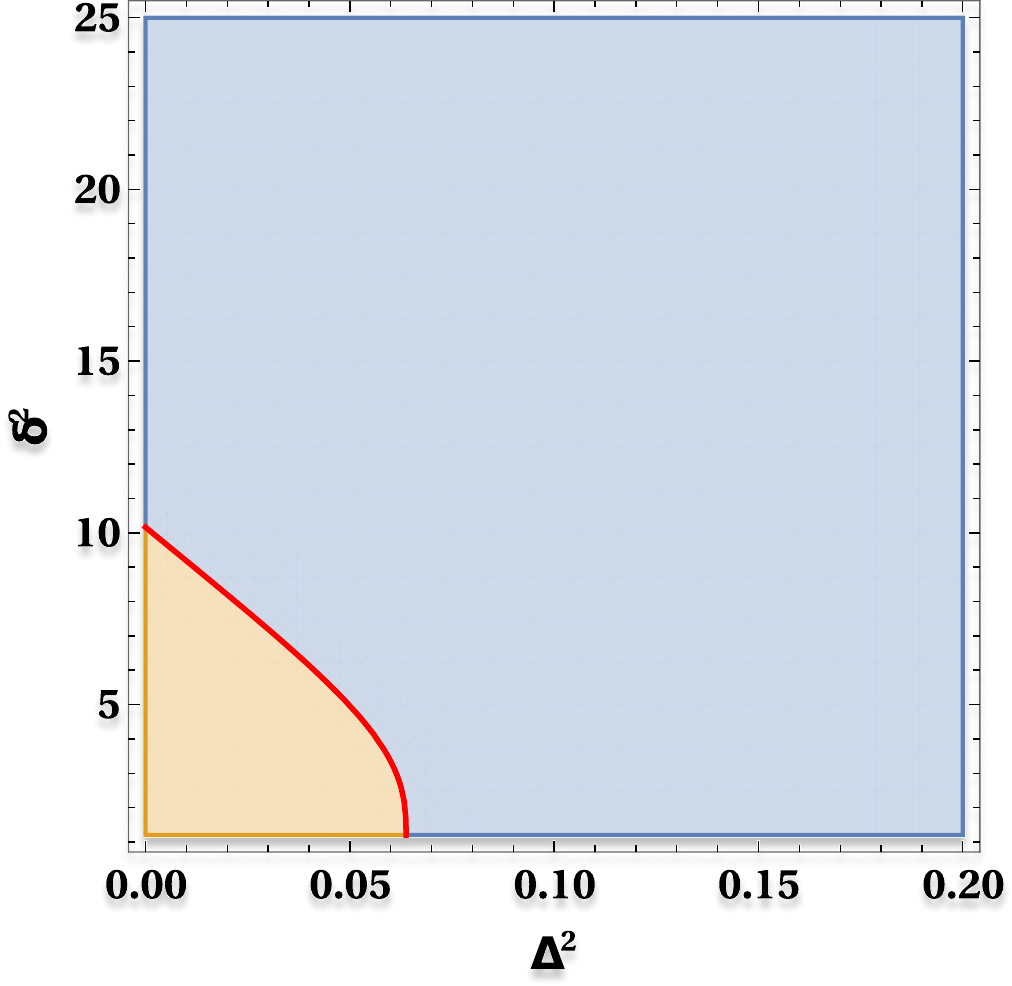}}
\subfloat[]{\includegraphics[width = 0.25 \textwidth, height=0.19\textwidth]{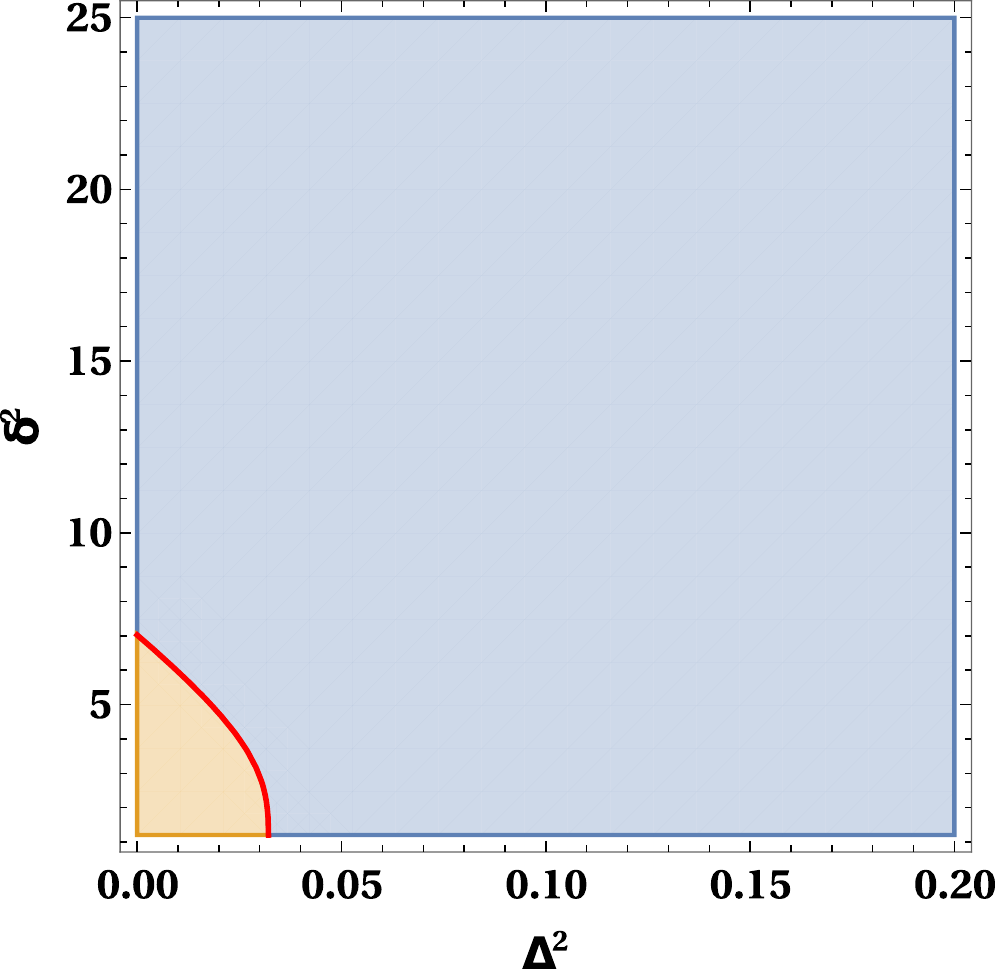}}
\caption{In the figure the orange shaded portion describes the region of $\Delta$ and $\delta$ for which $\mathcal{B}_{m}$ lies in quantum nonlocal regime with the red solid line separating it from the classical local realist regime ( represented by shaded sky portion). Each point on the red solid line is a representative of a transition point pair $(\delta_{n,m}^{c},\Delta_{n,m}^{c})$. Input settings $m$ = 2, 3, 4, and 5 are plotted in (a),(b), (c), and (d) respectively.}
\label{fig_bell3d}
\end{figure}
We numerically optimize the nonlocality witness $\mathcal{B}_{m}$ (in particular for $m$=2,3,4,5) involving the correlation function given by Eq. \eqref{eq:corfinal} and plot the region of $\Delta^{2}$ and $\delta^{2}$ for which $\mathcal{B}_{m}$ lies in the quantum regime with $n=5$. The reason for taking $n=5$ to observe the degree of coarsening at which the quantum-to-classical transition occurs as the degree of macroscopicity of the entangled state increases is that it is the lowest value for which every nonlocality witness is showing at least a nonzero amount of violation of local-realist bound. Then orange portions of the graphs are representative of the ranges of values of the degree of coarsening $\Delta$ and $\delta$ for which the macroscopic entangled state exhibits quantum nonlocality. The area of the orange region is a measure of the tensility of the \textit{quantumness} of the entangled state with $n=5$. Moreover, each point on the solid red curve defines a transition point pair $(\delta_{n,m}^{c},\Delta_{n,m}^{c})$ beyond (sky-colored region) which the state no longer shows nonlocality. It is also apparent from the observations of graphs in Fig.\ref{fig_bell3d} that instead of a uniform decrease of the orange shaded area, it shows a distinctive pattern as the measurement settings increase. Notably, when the number of settings is even there is a decrement in the quantum region. Conversely, in scenarios characterized by an odd number of settings an increment in the area of the aforementioned area is observed. In that sense, it follows the same trend as in the case of the coarsening of final measurement resolution.

From the above discussions, it is now evident that the symmetric nonlocality witness $\mathcal{B}_{m}$ fails to capture any delay in the quantum-to-classical transition for any arbitrary number of settings used to demonstrate nonlocality in terms of complete coarse-grained measurement. Thus, it is better to look for other nonclassicality witness instead of nonlocality. In this regard, we will now use quantum steering as a nonclassical resource and study its tensility for the macroscopic entangled states under complete coarse-grained measurement.
\section{Effect of complete coarse-grained measurement on steering}
The apparent inconsistency of the \textit{quantumness} exhibited in the form of nonlocality compels us to investigate a comparatively weaker form of bipartite quantum correlation namely quantum steering. Similar to the previous case we study the tensility of the quantum characteristics of the macroscopic entangled state under the complete coarse-grained measurement. In order to do this we introduce the n-settings linear steering inequality \cite{E.G.Cavalcanti} given by,
\begin{equation}\label{swit}
    \mathcal{S}_{m} \equiv \frac{1}{\sqrt{m}}\left|\sum_{i=1}^{m}\langle\hat{A}_{i}\otimes\hat{B}_{i}\rangle\right| \leq 1
\end{equation}
where we denote $\mathcal{S}_{m}$ as the steering witness with $m$ measurement settings per party in a bipartite scenario where  $\hat{A}_{i}$ and $\hat{B}_{i}$ with $i\in \{1,m\}$ are observables corresponding to the measurements performed by Alice and Bob respectively. If for a macroscopic entangled state $\ket{\Psi_{n}}$ and some choice of measurements of Alice and Bob, the joint statistics violate the above inequality we say that the state is steerable. When we consider the fuzzy version of the correlation functions similar to the nonlocality, at a certain degree of coarsening of measurement the statistics can no longer violate the inequality implying a quantum-to-classical transition of the macroscopic entangled state. However, in contrast to Bell nonlocality, in this case, unsteerability is referred to as the relevant notion of classicality.
\begin{figure}[h]
\centering
\subfloat[]{\includegraphics[width = 0.25 \textwidth, height=0.19\textwidth]{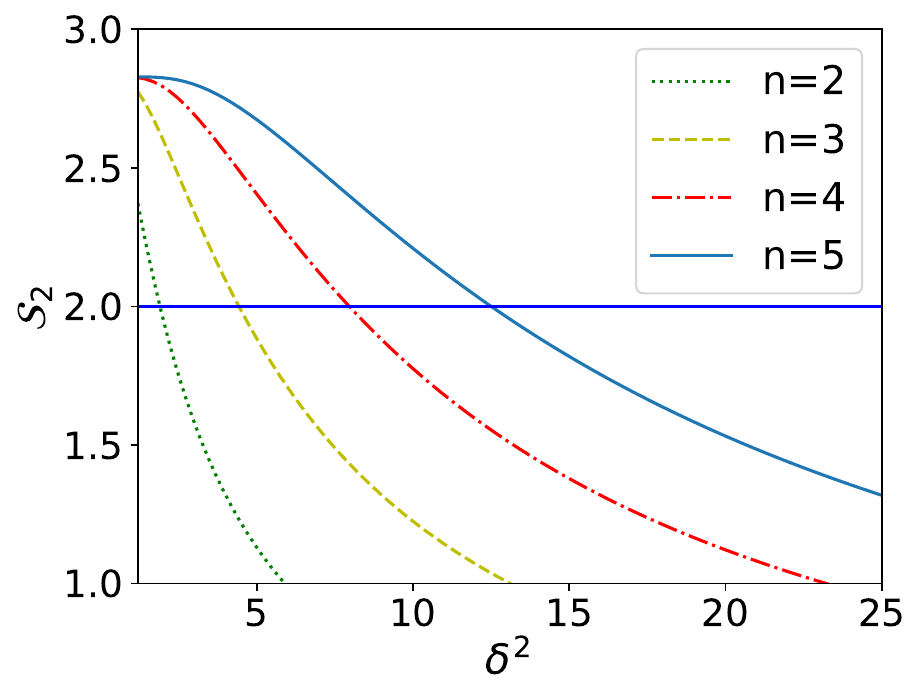}} 
\subfloat[]{\includegraphics[width = 0.25 \textwidth, height=0.19\textwidth]{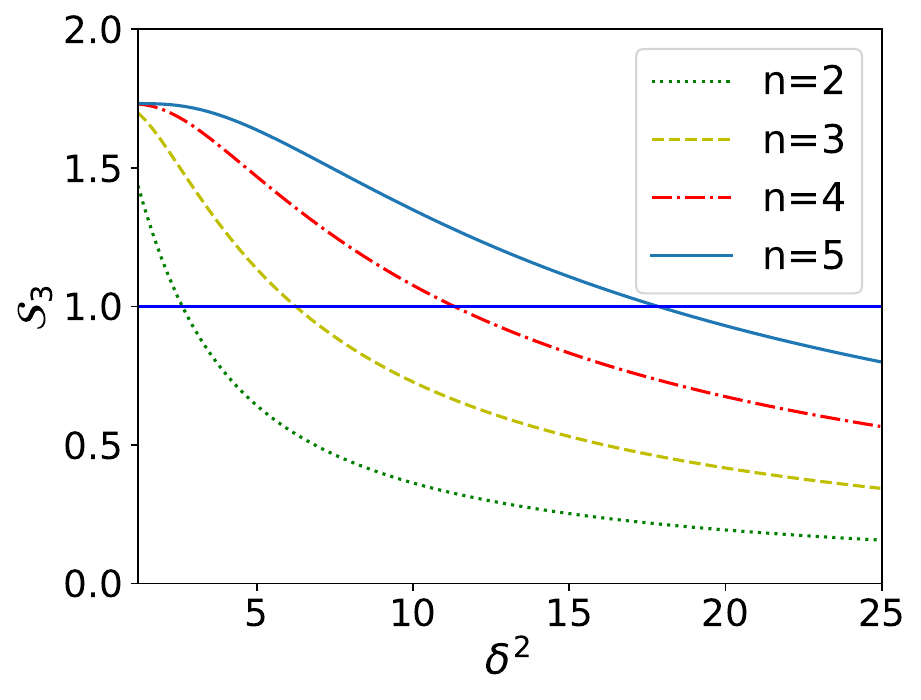}} \\
\subfloat[]{\includegraphics[width = 0.25 \textwidth, height=0.19\textwidth]{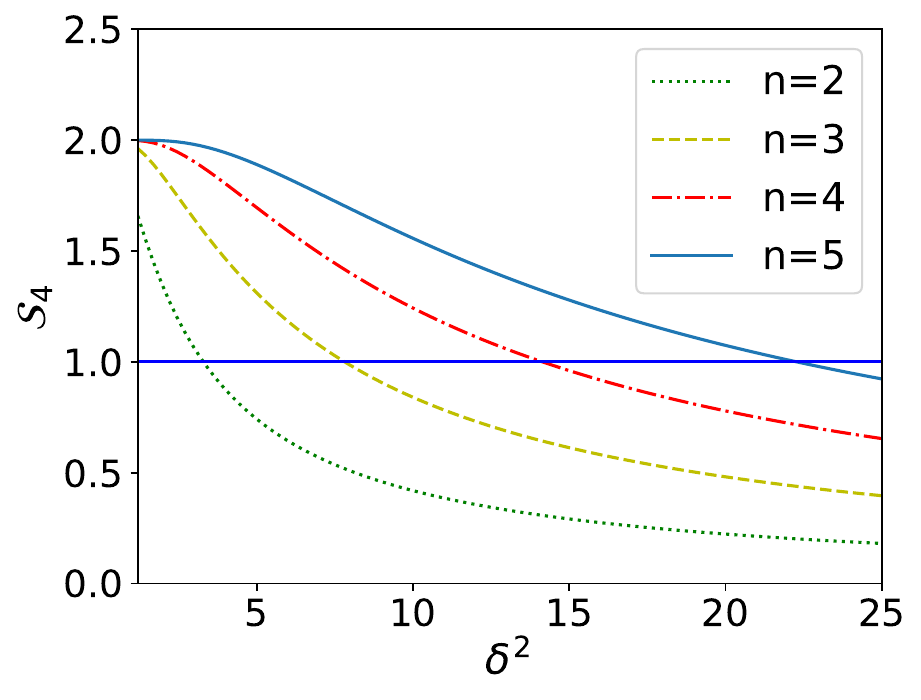}}
\subfloat[]{\includegraphics[width = 0.25 \textwidth, height=0.19\textwidth]{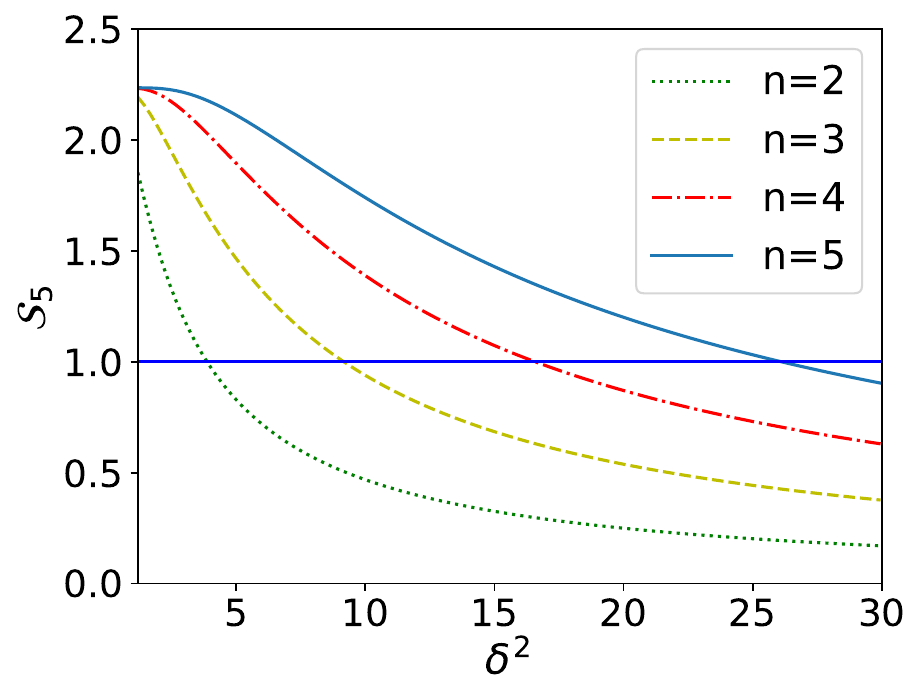}}
\caption{In this figure numerically optimized steering witness $\mathcal{S}_m$ has been plotted against the the variance $\delta^{2}$. The blue line represents the classical bound in each case. Input settings
$m$ = 2, 3, 4, and 5 are plotted in (a),(b), (c), and (d) respectively.}
\label{fig_str}
\end{figure}

 We begin with the situation where the fuzziness is involved only in the final resolution of measurement outcomes ($\delta > 1$) with perfect control on the measurement reference ($\Delta=0$). In such case the fuzzy version of the bipartite correlation function, $\langle\hat{A}_{i} \hat{B}_{i}\rangle_{\delta}= \langle\hat{A}_{\delta} (\phi_{i}) \otimes  \hat{B}_{\delta} (\phi_i)\rangle$, due to the measurements $A_{i}$ and $B_{i}$ performed respectively by Alice and Bob on their subsystem is of the same form as given in Eq.\eqref{eq9}. The steering witness $\mathcal{S}_{m}$ is thus calculated by substituting Eq.\eqref{eq9} into Eq. \eqref{swit} as a function of the degree of coarsening $\delta$. The numerically optimized value of $\mathcal{S}_{m}$ is plotted as a function of $\delta^{2}$ in Fig.\ref{fig_str} above. In contrast to the case where nonlocality is taken as a measure of \textit{quantumness}, here we observe that the effect of fuzziness on quantum steering exhibited by a macroscopic entangled state (with a certain degree of macroscopicity $n$) can be compensated by testing a steering witness with more number of measurement settings. More precisely, the critical value $\delta_{n,m}^{s}$ of the degree of coarsening at which a certain steering witness $\mathcal{S}_{m^{*}}$ stops violating Eq.\eqref{swit} can be increased by testing a steering witness $\mathcal{S}_{m}$ such that $m^{*}< m$. This trend is explicitly depicted in Fig.\ref{fig_str}, where different figures (Fig.\ref{fig_str}(a-d)) shows plots for different number of settings ($m$=2,3,4,5) in which each of them consists of plots of the steering witness 
 $\mathcal{S}_{m}$ as a function of $\delta^{2}$ for different values of $n$.

\begin{figure}[H]
\centering
\subfloat[]{\includegraphics[width = 0.25 \textwidth, height=0.19\textwidth]{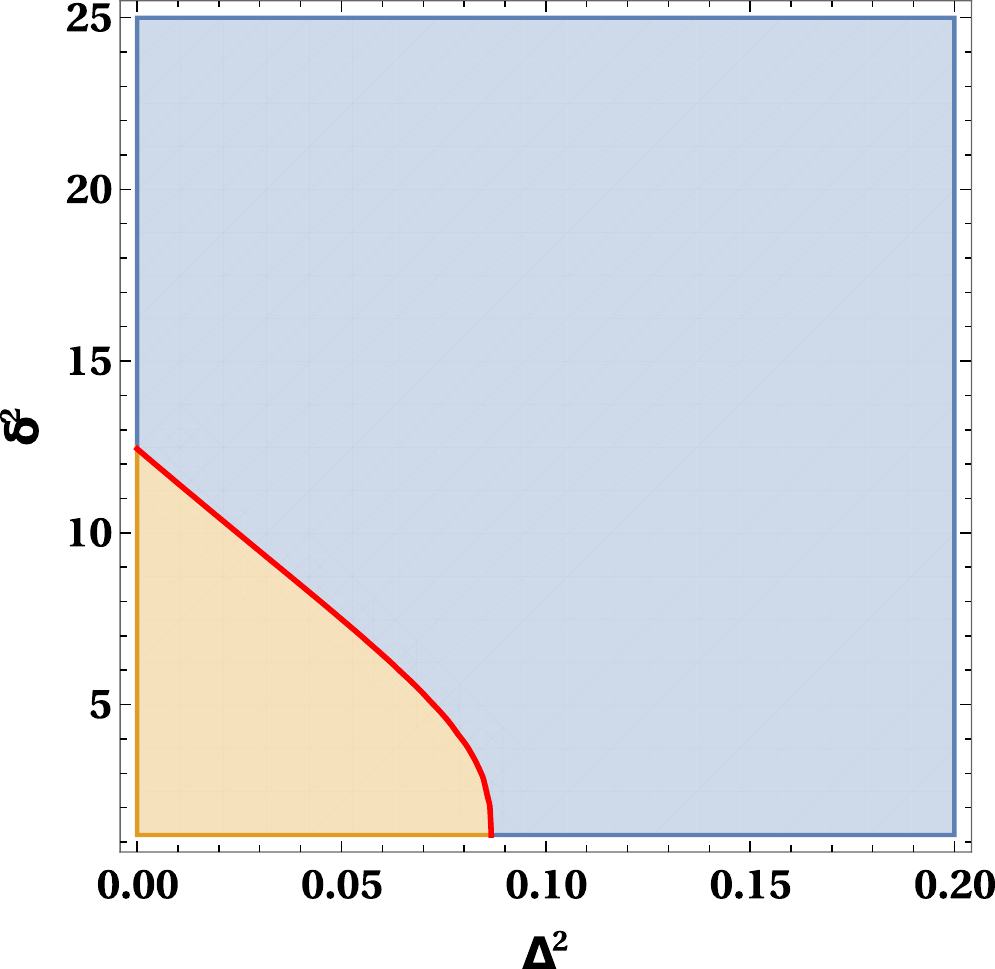}} 
\subfloat[]{\includegraphics[width = 0.25 \textwidth, height=0.19\textwidth]{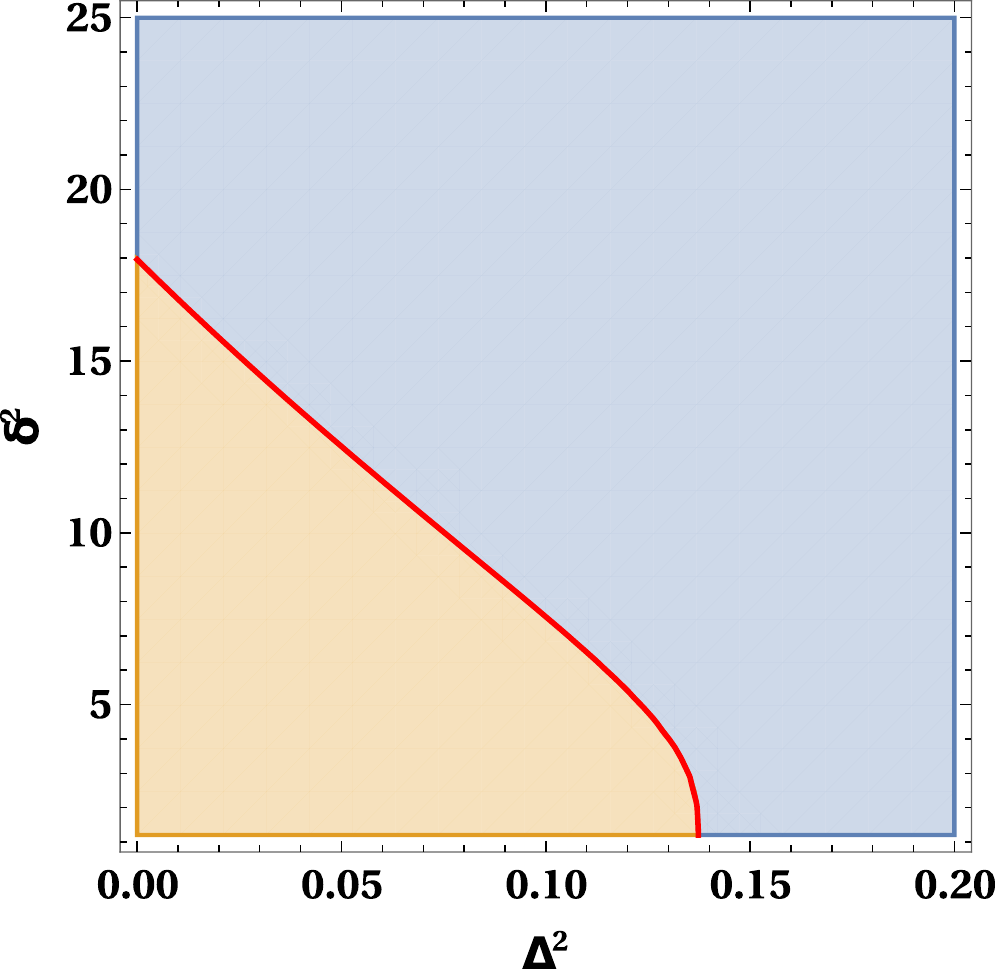}} \\
\subfloat[]{\includegraphics[width = 0.25 \textwidth, height=0.19\textwidth]{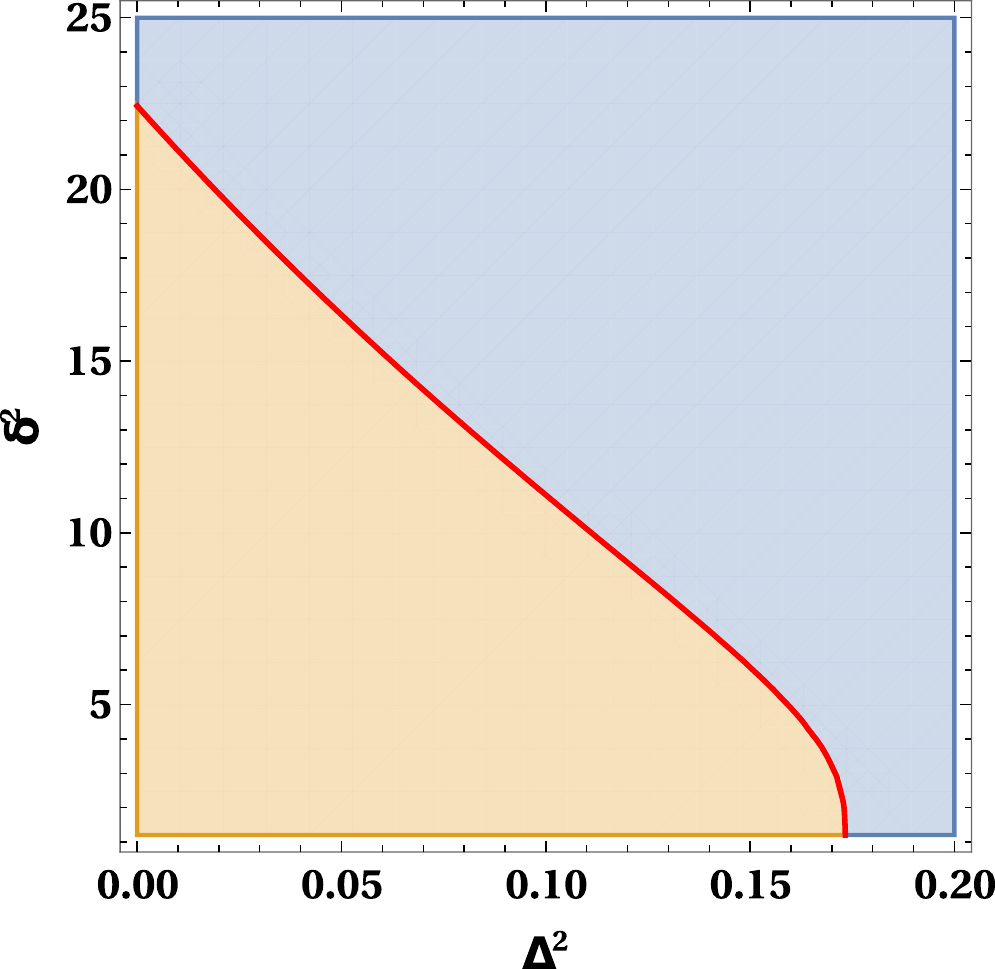}}
\subfloat[]{\includegraphics[width = 0.25 \textwidth, height=0.19\textwidth]{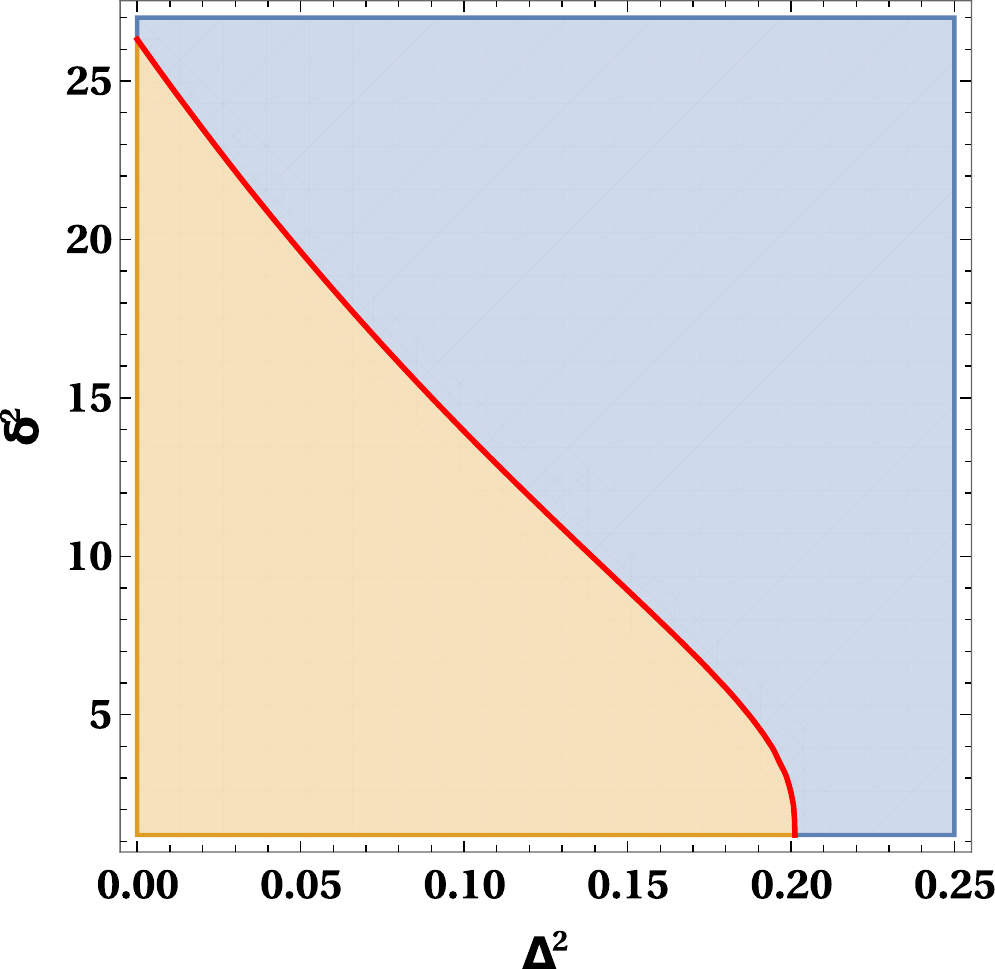}}
\caption{In the figure, the orange shaded area shows where $\mathcal{S}_{m}$ enters the quantum nonlocal zone, as separated by the solid red line from the classical local realist regime (shown in light blue). Each point on the red line represents a pair of transition points $(\delta_{n,m}^{s},\Delta_{n,m}^{s})$. Panels (a), (b), (c), and (d) correspond to input settings with $m$ equal to 2, 3, 4, and 5, respectively. }
\label{fig_ccs}
\end{figure}

Let us now consider the most general scenario where the effect of fuzziness on both the resolution of final measurement outcomes as well as measurement reference is nonzero.  For the macroscopic entangled state given by Eq.\eqref{eq3}, the joint correlation function $\langle\hat{A}_{i} \hat{B}_{j}\rangle_{\delta, \Delta}$ is given by Eq.\eqref{eqccg}. This correlation function is employed to calculate the steering witness $\mathcal{S}_{m}$ for given n and m. The effect of the complete coarse-grained measurement on the amount of \textit{quantumness} revealed in a macroscopic entangled state is represented through Fig.\ref{fig_ccs} for different steering witness $\mathcal{S}_{m}$ with $(m=2,3,4,5)$ by taking $n$ = 5 as the degree of macroscopicity as an example. In each of the graphs, the orange shaded region showcases the range of values of $\Delta^{2}$ and $\delta^{2}$ for which the state exhibits quantum steering by violating the inequality given in Eq.\eqref{swit}. The area of each of the plots thus reveals the tensility of quantum steering with respect to the strength of coarsening. Any transition point $(\Delta^{2}, \delta^{2})$ that lies in the sky-colored shaded region gives the pair of values $(\Delta, \delta)$ for which inequality Eq.\eqref{swit} is satisfied indicating the absence of \textit{quantumness} (in the form of quantum steering) in the macroscopic entangled state. Thus, the points on the solid red curve give the trade-off between two degrees of coarsening parameters $\delta_{n,m}^{s}$ and $\Delta_{n,m}^{s}$ for which quantum-to-classical transition occurs. It is now clear from the Fig.\ref{fig_ccs} (a-d) that the degree of coarsening at which the quantum-to-classical transition occurs in terms of a  steering witness $\mathcal{S}_{m}$ can be consistently pushed forward by testing another witness of increased number of settings. Thus the effect of coarsening on the amount of steering certain macroscopic quantum state exhibits can be compensated by increasing the number of settings of the steering witness that is employed to test the steering.

\begin{figure}[H]
  \centering
  \includegraphics[scale=0.45]{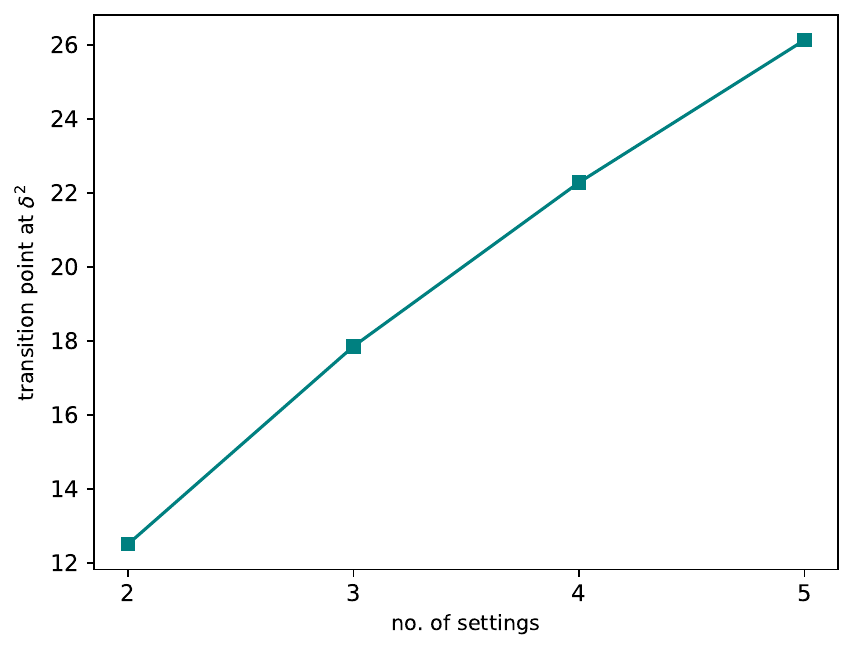}
\caption{ In the figure the Value of $\delta^{2}$ corresponding to the transition points (for quantum steering) is plotted against the numbers of measurement settings with the degree of macroscopicity $n=5$ of the entangled state. }
  \label{consistency}  
\end{figure}
\section{Effect of white noise}
So far we have only discussed the tensility of the \textit{quantumness} of a pure macroscopic entangled state. However, as several experimental situations demand, it is essential to investigate the robustness of the nonclassicality against the noise introduced in the source.  In this section, we will study the robustness of quantum correlations (Bell's nonlocality and quantum steering) under a complete coarse-grained measurement scenario in terms of mixed entangled states. In order to do so let us first consider a system consisting of pure macroscopic maximally entangled state supplemented with white noise ( we call it macroscopic Werner state) as,
\begin{equation}\label{weq1}
   \rho_{mix} = p\ket{\Psi_{n}}\bra{\Psi_{n}}+\left(\frac{1-p}{4}\right)\mathbb{1}.
\end{equation}
It is already mentioned that for $p>\frac{1}{\sqrt{2}}$ the Werner state shows Bell nonlocality, quantum steering for $p>\frac{1}{2}$  and entanglement for $p>\frac{1}{3}$. Therefore, in the range $\frac{1}{3}<p \leq \frac{1}{2}$, this state is only entangled and exhibits neither Bell nonlocality nor quantum steering. It is steerable and entangled but not Bell nonlocal for $\frac{1}{2}<p \leq \frac{1}{\sqrt{2}}$, while for $p > \frac{1}{\sqrt{2}}$, it exhibits Bell nonlocality along with quantum steering and entanglement. In order to find the value of the quantum witnesses we have to first obtain the joint bipartite correlation functions  
$\langle \hat{A}_{i},\hat{B}_{j} \rangle_{\delta, \Delta}^{p} = \left\langle \hat{A}_{\delta,\Delta}(\phi_{i})\otimes\hat{B}_{\delta,\Delta}(\phi_{j})\right\rangle^{p}$ for the state $\rho_{mix}$. As in the previous cases, here we will first introduce the fuzziness solely in the resolution of the final measurement, then in the measurement reference, and finally in both of these and then study the quantum witnesses for both nonlocality and steering.

Let us first consider the case where no fuzziness has been incorporated in the unitary transformation which implies $\Delta=0$ and the final detection of the measurement is inaccurate ($\delta>0$). The correlation function can be written as,
\begin{multline}\label{weq2}
    \langle\hat{A}_{i} \hat{B}_{j}\rangle_{\delta}^{p} = \frac{p}{2}[ \mathcal{Q}_{\delta}(n,\phi_{a})\hspace{0.1cm}\mathcal{Q}_{\delta}(-n,\phi_{b}) \\ + \mathcal{Q}_{\delta}(-n,\phi_{a})\hspace{0.1cm}\mathcal{Q}_{\delta}(n,\phi_{b}) + 2\mathcal{R}_{\delta}(n,\phi_{a})\hspace{0.1cm}\mathcal{R}_{\delta}(n,\phi_{b}) ]\\ + \left(\frac{1-p}{4}\right)[\mathcal{Q}_{\delta}(n,\phi_{a})\mathcal{Q}_{\delta}(n,\phi_{b})+\mathcal{Q}_{\delta}(n,\phi_{a})\mathcal{Q}_{\delta}(-n,\phi_{b})\\+\mathcal{Q}_{\delta}(-n,\phi_{a})\mathcal{Q}_{\delta}(n,\phi_{b})+\mathcal{Q}_{\delta}(-n,\phi_{a})\mathcal{Q}_{\delta}(-n,\phi_{b})],
\end{multline}
where $\mathcal{Q}_{\delta}(n,\phi) $ and $\mathcal{R}_{\delta}(n,\phi)$ is defined earlier as given in Eq.\eqref{eq9}.\\

We have shown in the preceding sections that when $m=2$, the degree of coarsening for which a pure macroscopic entangled state undergoes quantum-to-classical transition is the same ($\delta_{n,2}^{c}=\delta_{n,2}^{s}$) for both Bell's nonlocality and quantum steering. Surprisingly, for a particular mixed state with fixed value of $p$, we found that even in the case of $m=2$ the quantum-to-classical transition points are different, i.e., $\delta_{n,2}^{c} \neq \delta_{n,2}^{s}$. We also found as the value of visibility $p$ approaches unity, $\delta_{n,m}^{c}$ and $ \delta_{n,m}^{s}$ becomes closer to one another and become equal for $p=1$. As of our knowledge this particular behaviour of a macroscopic mixed entangled state under coarse-grained measurement has not been noticed in previous studies. In order to illustrate this trend for $m = 2$, an example of a set of transition points corresponding to different values of $p$ for a macroscopic entangled state with $n = 5$ is shown in Tab. \ref{tab1}.  As the form of the correlation function is same for both the cases, it is evident that the form of the inequalities are viable for such differences. One can find some other form of symmetric steering inequalities for which this disparity evaporates. For any other value of $m$ with fixed $\delta$ we observe monotonic decrement of the value of both the quantum witnesses $\mathcal{B}_{m}$ and $\mathcal{S}_{m}$ with an increase in the visibility $p$.\\

We get more strict constraint on the \textit{quantumness} of a mixed macroscopic entangled state by considering a complete coarse-grained measurement performed by the respective parties. In this case the joint correlation function takes the form,
\begin{multline}\label{weq4}
    \langle\hat{A}_{i} \hat{B}_{j}\rangle_{\delta,\Delta}^{p} = \int_{-\infty}^{\infty}\int_{-\infty}^{\infty}d\phi_{a}d\phi_{b}P_{\Delta}(\phi_{a}-\theta_{a})P_{\Delta}(\phi_{b}-\theta_{b})\\ \text{Tr}[(U^\dagger(\phi_{a})F_{\delta}U(\phi_{a}))\otimes(U^\dagger(\phi_{b})F_{\delta}U(\phi_{b}))\rho_{mix} ] \\
    = \int_{-\infty}^{\infty}\int_{-\infty}^{\infty}d\phi_{a}d\phi_{b}P_{\Delta}(\phi_{a}-\theta_{a})P_{\Delta}(\phi_{b}-\theta_{b})\\ \left[pM_{\delta}(n,\phi) + \left(\frac{1-p}{4}\right)T_{\delta}(n,\phi)\right]
\end{multline}
where, $M_{\delta}(n,\phi)=\mathcal{Q}_{\delta}(n,\phi_{a})\mathcal{Q}_{\delta}(-n,\phi_{b}) + \mathcal{Q}_{\delta}(-n,\phi_{a})\mathcal{Q}_{\delta}(n,\phi_{b}) + 2\mathcal{R}_{\delta}(n,\phi_{a})\mathcal{R}_{\delta}(n,\phi_{b})$\\ and $T_{\delta}(n,\phi)=\mathcal{Q}_{\delta}(n,\phi_{a})\mathcal{Q}_{\delta}(n,\phi_{b})+\mathcal{Q}_{\delta}(n,\phi_{a})\mathcal{Q}_{\delta}(-n,\phi_{b})+\mathcal{Q}_{\delta}(-n,\phi_{a})\mathcal{Q}_{\delta}(n,\phi_{b})+\mathcal{Q}_{\delta}(-n,\phi_{a})\mathcal{Q}_{\delta}(-n,\phi_{b}).$\\\\
The visibility $p$ for which a particular mixed macroscopic entangled state makes quantum-to-classical transition for different values of $\delta$ and $\Delta$ is calculated by putting Eq.\eqref{weq4} into the expression of $\mathcal{B}_{m}$ and $\mathcal{S}_m$ and noting its classical bound. However, in order to avoid clumsiness we  are not including (except for $m=2$) numerical values of visibilities for different measurement settings and different degrees of coarsening. It is evident from Tab. \ref{tab1} that for $m=2$ unlike $\delta^{2}$ that takes different value corresponding to the transition points for Bell nonlocality and quantum steering even with fixed value of $p$, the value of $\Delta^{2}$ corresponding to the transition points is the same and is independent of which of these two is taken as a measure of quantumness. 
\begin{table}[]
\addtolength{\tabcolsep}{5.4pt}
\begin{tabular}{ccccc}
\hline\hline
\multicolumn{1}{l}{\multirow{2}{*}{\textbf{Visibility}}} & \multicolumn{2}{c}{\textbf{Nonlocality}}    & \multicolumn{2}{c}{\textbf{Steering}}       \\ \cline{2-5} 
\multicolumn{1}{l}{}                                     & $\delta^2(\Delta=0)$ & $\Delta^2(\delta=0)$ & $\delta^2(\Delta=0)$ & $\Delta^2(\delta=0)$ \\ \hline
p=0.85 & 8.29    & 0.046  & 8.94  & 0.046  \\
p=0.80  & 6.72    & 0.0308 & 7.615 & 0.0308 \\
p=0.75 & 4.81042 & 0.0147 & 6.137 & 0.0147 \\ \hline\hline
\end{tabular}
\caption{Table shows transition points in-term of variance $(\delta^{2}, \Delta^{2})$, for $m=2$ with varying degrees of visibility $p$.  }
\label{tab1}
\end{table}

\section{Concluding remarks}
The quantum-to-classical transition of a macroscopic entangled state \cite{H.Jeong} due to different forms of inefficiency of measuring apparatus is investigated. In particular, we focus on the interplay between the degree of coarsening of measurement and the number of measurement settings used in revealing the \textit{quantumness} (nonlocality and quantum steering) of the macroscopic entangled state with certain degree of macroscopicity. As the degree of coarsening of the measurement increases, the non-classical behaviour begins to fade up and eventually disappear as it crosses a certain threshold value. We found that such decline of \textit{quantumness} of the system depends on the number of input settings of the particular witness employed to reveal the nonclassicality. Most importantly, we have shown particular instances for which the effect of coarsening can be compensated by increasing the number of measurement settings.

We have first considered the violation of a family of Bell inequalities (characterized by different number of measurement settings) by a macroscopic entangled state under coarsening of both final measurement resolution as well as the measurement reference. This degree of fuzziness of such coarse-grained measurement is one of the eligible candidate on which the quantum-to-classical transition depends. We have first studied the tensility of the nonlocality exhibited by the macroscopic entangled states under varying strength (degree) of coarsening. It is shown that similar to ref. \cite{H.Jeong}, that only considered a two $(m=2)$ measurement settings symmetric nonlocal witness, for $m \geq 2$ with any particular nonlocal witness quantum to classical transition occurs at a greater degree of coarsening as we increase the macroscopicity of the entanglement. However a disparity in transition points is seen corresponding to even and odd number of settings. Notably, when the number of settings is even, the value of the degree of coarsening corresponding to the transition point decreases as number of measurement settings increases. Conversely, for the odd number of settings an increase of the number of measurements leads to an increment of the aforementioned value. Moreover, as one increases the number of settings substantially, the transition points corresponding to even and odd number of settings eventually converges and the quantum-to-classical transition becomes independent of the number of measurement settings. This disparity restricts this particular set of symmetric Bell inequalities \cite{N.Gisin1999} from establishing a consistent relationship between the trio namely, the macroscopicity of the entanglement, number of measurement settings of the nonlocality witness, and the degree of coarsening of measurements.

Against the backdrop, by considering unsteerability as a form of classicality, we finally employed a family of linear steering inequalities to witness the quantum-to-classical transition. It is clearly depicted that as the degree of macroscopicity increases, the quantum states becomes more robust to the fuzziness of measurement. In terms of the the nature of the trade-off between the macroscopicity and the amount of \textit{quantumness}, quantum steering resembles with the nonlocality as discussed earlier. What singles out quantum steering from nonlocality is the number of measurement settings one uses to witness the quantumness of the system. It is shown that with the increase in the number of input measurement settings of the steering witness, the degree of coarsening corresponding to the transition points increases. Moreover, in comparison to nonlocality, for a particular entangled state and fixed number of measurement settings the steering witness is shown to be more robust than the nonlocality witness. Thus not only with respect to the mixedness of the entangled state, but also in terms of coarsening of measurements, steering remains as the weaker correlation compared to the nonlocality.

While concluding we note that the disparity between the set of nonlocality witnesses with even and odd number of settings in extracting the nonlocality exhibited by the macroscopic entangled state can be eliminated by considering asymmetric nonlocality witness \cite{GisinN}. It will also be interesting to study effects of the particular aspects of coarsening of measurements discussed in this paper on correlations like macroscopic nonlocality \cite{M.Gallego, M.Navascués} as well as macroscopic contextuality \cite{J.Henson}, in order to explore more finer details about the quantum-to-classical transition. This calls for further study.

\section{Acknowledgement}
LPN and TG acknowledge the hospitality of IISER Kolkata, SM acknowledges the support from IISER Kolkata institute postdoctoral fellowship. PKP acknowledge the support from DST, India, through Grant
No. DST/ICPS/QuST/Theme-1/2019/2020-21/01.

\end{document}